\documentclass[a4paper,fleqn,10pt]{article}
\pdfoutput=1
\usepackage{amsmath}
\usepackage{amssymb}
\usepackage{array}
\usepackage{calc}
\usepackage{longtable}
\usepackage{multirow}
\usepackage{slashed}
\usepackage{pstricks}
\usepackage{graphicx}
\usepackage{xspace}
\usepackage{siunitx}[=v2]
\usepackage{booktabs}
\usepackage{tikz}
\usepackage{textcomp}

\numberwithin{equation}{section}
\usepackage{cite}
\usepackage[pdfborder={0 0 0}]{hyperref}
\usepackage[format=hang,labelfont=bf,hypcap=true]{caption}
\usepackage{subcaption}

\usepackage{listings}
\newsavebox{\myboxone}
\newsavebox{\myboxtwo}
\usepackage{mathabx}
\usepackage{sectsty}
\usepackage{enumitem}
\allsectionsfont{\sffamily}
\subsubsectionfont{\mdseries\itshape\large}
\makeatletter
\DeclareRobustCommand*{\bfseries}{%
  \not@math@alphabet\bfseries\mathbf
  \fontseries\bfdefault\selectfont
  \boldmath
}
\makeatother
\let\spreprint\empty
\newcommand{\preprint}[1]{\def\spreprint{\protect#1}}
\let\sinstitute\empty
\newcommand{\institute}[1]{\def\sinstitute{\protect#1}}
\makeatletter
\renewcommand{\maketitle}{\begingroup
  \null\thispagestyle{empty}%
    \ifx\spreprint\empty
      \vskip 5ex
    \else
      \flushright\large\spreprint\vskip 10ex
    \fi
    \vskip 5ex
    \flushleft
      {\sffamily\bfseries\huge\@title}\vskip 6ex
      \@author\vskip 2ex
      \ifx\sinstitute\empty
      \else
        {\small\sinstitute}
      \fi
    \vskip 5ex
  \endgroup
}
\makeatother
\renewenvironment{abstract}{\begin{center}
  {\large\sffamily\bfseries Abstract: }
  \begin{minipage}[t]{0.75\textwidth}
}{\end{minipage}\end{center}\vskip 10ex}

\numberwithin{equation}{section}
\allowdisplaybreaks[2]
\newcommand{\eg}{\textit{e.g.\@}\xspace}
\newcommand{\ie}{\textit{i.e.\@}\xspace}

\newcommand{\see}{\textit{see\@}\xspace}

\newcommand{\SMCatNLO}{S--M\protect\scalebox{0.8}{C}@N\protect\scalebox{0.8}{LO}\xspace}
\newcommand{\LCMCatNLO}{$\langle\text{LC}\rangle$--M\protect\scalebox{0.8}{C}@N\protect\scalebox{0.8}{LO}\xspace}
\newcommand{\LCMCatNLOCSS}{$\langle\text{LC}\rangle$--M\protect\scalebox{0.8}{C}@N\protect\scalebox{0.8}{LO}--\CSS{}\xspace}
\newcommand{\MCatNLO}{M\protect\scalebox{0.8}{C}@N\protect\scalebox{0.8}{LO}\xspace}

\newcommand{\MEPSatNLO}{M\scalebox{0.8}{E}P\scalebox{0.8}{S}@N\scalebox{0.8}{LO}\xspace}
\newcommand{\CKKW}{CKKW\xspace}

\newcommand{\LHAPDF}{L\protect\scalebox{0.8}{HAPDF}\xspace}

\newcommand{\BlackHat}{B\protect\scalebox{0.8}{LACK}H\protect\scalebox{0.8}{AT}\xspace}

\newcommand{\MCFM}{M\protect\scalebox{0.8}{CFM}\xspace}

\newcommand{\OpenLoops}{O\protect\scalebox{0.8}{PEN}\-L\protect\scalebox{0.8}{OOPS}\xspace}

\newcommand{\Sherpa}{S\protect\scalebox{0.8}{HERPA}\xspace}
\newcommand{\Comix}{C\protect\scalebox{0.8}{OMIX}\xspace}

\newcommand{\Amegic}{A\protect\scalebox{0.8}{MEGIC}\xspace}
\newcommand{\CSS}{C\protect\scalebox{0.8}{SS}\xspace}

\long\def\symbolfootnote[#1]#2{\begingroup%
\def\thefootnote{\fnsymbol{footnote}}\footnote[#1]{#2}\endgroup}

\newcommand{\done}{{\mathrm{d}}}

\newcommand{\alphaS}{\alpha_s}

\newcommand{\bea}{\begin{eqnarray}}
\newcommand{\eea}{\end{eqnarray}}
\newcommand{\bi}{\begin{itemize}}
\newcommand{\ei}{\end{itemize}}

\newcommand{\muR}{\ensuremath{\mu_\mathrm{R}}\xspace}
\newcommand{\muF}{\ensuremath{\mu_\mathrm{F}}\xspace}
\newcommand{\muRF}{\ensuremath{\mu_\mathrm{R,F}}\xspace}

\newcommand{\EWvirt}{\ensuremath{\text{EW}_\text{virt}}}

\xspace
\xspace
\xspace
\xspace
\xspace
\xspace
\xspace
\xspace
\xspace
\xspace
\xspace
\xspace
\xspace
\xspace

\newcommand{\eeproc}{\ensuremath{pp \to e^+e^- + 0,1,2j\text{@NLO}+3,4,5j\text{@LO}}}
\newcommand{\ttproc}{\ensuremath{pp \to t\bar{t} + 0,1j\text{@NLO}+2,3,4j\text{@LO}}}

\newlist{myitemize}{itemize}{3}
\setlist[myitemize]{leftmargin=14em}

\newcolumntype{C}{>{\centering\arraybackslash}p{0.14\textwidth}}

\newlength{\unitcharwidth}
\newcommand{\NLOCSSPSMODEi}[1]{\text{\texttt{NLO\_CSS\_PSMODE=#1}}}
\newcommand{\UnwgtGenX}[1]{\text{\texttt{Pilot\_Loop\_Generator=#1}}}

\newcommand{\TeV}{\ensuremath{\text{Te\kern -0.1em V}\xspace}}
\newcommand{\GeV}{\ensuremath{\text{Ge\kern -0.1em V}\xspace}}
\newcommand{\MeV}{\ensuremath{\text{Me\kern -0.1em V}\xspace}}
\newcommand{\keV}{\ensuremath{\text{ke\kern -0.1em V}\xspace}}
\newcommand{\eV}{\ensuremath{\text{e\kern -0.1em V}\xspace}}

\hypersetup{
  pdfauthor={Enrico Bothmann, Andy Buckley, Ilektra A. Christidi, Christian Gutschow, Stefan Hoeche, Max Knobbe, Marek Schoenherr},
  pdftitle={Accelerating LHC event generation with simplified pilot runs and fast PDFs}
}
\preprint{FERMILAB-PUB-22-462-T\\IPPP/22/39\\MCnet-22-17}
\author{Enrico Bothmann$^1$, Andy Buckley$^2$, Ilektra A. Christidi$^3$, Christian G{\"u}tschow$^4$,\\
  Stefan H{\"o}che$^5$, Max Knobbe$^{1,2}$, Tim~Martin$^6$, Marek Sch{\"o}nherr$^7$}
\title{Accelerating LHC event generation\\[1mm] with simplified pilot runs and fast PDFs}
\institute{
  $^1$Institut f{\"u}r Theoretische Physik, Georg-August-Universit{\"a}t G{\"o}ttingen, 37077 G{\"o}ttingen, Germany\\
  $^2$School of Physics \& Astronomy, University of Glasgow, Glasgow, G12 8QQ, UK\\
  $^3$Centre for Advanced Research Computing, University College London, Gower Street, London, WC1E 6BT, UK\\
  $^4$Department of Physics and Astronomy, University College London, Gower Street, London, WC1E 6BT, UK\\
  $^5$Theory Division, Fermi National Accelerator Laboratory, Batavia, IL, 60510, USA\\
  $^6$Department of Physics, University of Warwick, Coventry, CV4 7AL, UK\\
  $^7$Institute for Particle Physics Phenomenology, Department of Physics, Durham University, Durham, DH1 3LE, UK\\
}

\begin{document}
\vspace*{10mm}
\maketitle
\begin{abstract}
  Poor computing efficiency of precision event generators for LHC physics
  has become a bottleneck for Monte-Carlo event simulation campaigns.
  We provide solutions to this problem by focusing on two major components
  of general-purpose event generators: The PDF evaluator and the matrix-element
  generator. For a typical production setup in the ATLAS experiment, we show
  that the two can consume about 80\% of the total runtime.
  Using NLO simulations of $pp\to\ell^+\ell^-+\text{jets}$ and
  $pp\to t\bar{t}+\text{jets}$ as an example, we demonstrate that the
  computing footprint of \LHAPDF and \Sherpa can be reduced by factors
  of order 10, while maintaining the formal accuracy of the event sample.
  The improved codes are made publicly available.
\end{abstract}


\section{Introduction}
\label{sec:intro}
Particle colliders have long dominated efforts to experimentally probe
fundamental interactions at the energy frontier.  They enable access to the
highest energy scales in human-made experiments, at high collision rates and
in controlled conditions, allowing a systematic investigation of the most basic
laws of physics. Event-generator programs have come to play a crucial role in such
experiments, starting with the use of early event generators such as
JETSET~\cite{Sjostrand:1982fn} and HERWIG~\cite{Webber:1983if} in the discovery
of the gluon at the PETRA facility in 1979.

Today, with the Large Hadron Collider (LHC) having operated successfully for
over a decade at nearly 1000 times the energy of PETRA, event generators are an
ever more important component of the software stack needed to extract
fundamental physics parameters from experimental
data~\cite{Buckley:2011ms,Campbell:2022qmc}. Most experimental measurements rely
on their precise modelling of complete particle-level events on which a detailed
detector simulation can be applied. The experimental demands on these tools
continue to grow: the precision targets of the high-luminosity LHC (HL-LHC)
~\cite{ZurbanoFernandez:2020cco} will
require both high theoretical precision and large statistical accuracy,
presenting major challenges for the currently available generator codes. With
much of the development during the past decades having focused on improvements
in theoretical precision---in terms of the formal accuracy of the elements
of the calculation---their computing performance has become a major
concern~\cite{Buckley:2019wov,HSFPhysicsEventGeneratorWG:2020gxw,
  HSFPhysicsEventGeneratorWG:2021xti,HEPSoftwareFoundation:2020daq}.

Event generators are constructed in a modular fashion, which is inspired by the
description of the collision events in terms of different QCD dynamics at
different energy scales. At the highest scales, computations can be carried out
using amplitudes calculated in QCD perturbation theory. These calculations have
been largely automated in matrix-element generators, both at
leading~\cite{Krauss:2001iv,Mangano:2002ea,Cafarella:2007pc,
  Gleisberg:2008fv,Alwall:2011uj}, and at next-to-leading~\cite{
  Berger:2008sj,Cullen:2011ac,Bevilacqua:2011xh,Cascioli:2011va,Alwall:2014hca,Actis:2016mpe}
orders in the strong coupling constant, $\alphaS$. Matrix-element generators
perform the dual tasks of computing scattering matrix elements fully
differentially in the particle momenta, as well as integrating these
differential functions over the multi-particle phase space using Monte Carlo
(MC) methods.

In principle, such calculations can be carried out for an arbitrary number of
final-state particles; in practice, the tractable multiplicities are very
limited. The presence of quantum interference effects in the matrix elements
induces an exponential scaling of computation complexity with the number of
final-state particles. This problem is exacerbated further by the rise of
automatically calculated next-to-leading order (NLO) matrix elements in the QCD
and electroweak (EW) couplings, which not only have a higher intrinsic cost from
more complex expressions, but are also more difficult to efficiently sample in
phase-space, and introduce potentially negative event weights which reduce the
statistical power of the resulting event samples. While theoretical work
progresses on these problems, e.g.\ by the introduction of rejection sampling using
neural network event-weight estimates~\cite{Danziger:2021eeg}, modified
parton-shower matching schemes~\cite{Frederix:2020trv,Danziger:2021xvr} and
resampling techniques~\cite{Andersen:2020sjs,Nachman:2020fff}, the net effect
remains that precision MC event generation comes at a computational cost far
higher than in previous simulation campaigns. Indeed, it already accounts for a
significant fraction of the total LHC computing
budget~\cite{Hoche:2019flt,HEPSoftwareFoundation:2020daq}, and there is a real
risk that the physics achievable with data from the high-luminosity runs
of the LHC will be limited
by the size of MC event samples that can be generated within fixed computing
budgets. It is therefore crucial that dedicated attention is paid to issues of
computational efficiency.

In this article, we focus on computational strategies to improve the performance
of particle-level MC event generator programs, as used to produce
large high-precision simulated event samples at the LHC. While the
strategies and observations are of a general nature, we focus our
attention on concrete implementations in the \Sherpa event
generator~\cite{Sherpa:2019gpd} and the
\LHAPDF library for parton distribution function (PDF)
evaluation~\cite{Buckley:2014ana}. Collectively, this effort is aimed at solving
the current computational bottlenecks in LHC high-precision event generation.
Using generator settings for standard-candle processes from the ATLAS
experiment~\cite{ATLAS:2021yza} as a baseline, we discuss timing improvements
related to PDF-uncertainty evaluation and for event generation more
generally. Overall, our new algorithms provide speedups of a factor
of up to 15 for the most time-consuming simulations in typical configurations, 
in time for the LHC Run-3 event-generation campaigns.

This manuscript is structured as follows: Section~\ref{sec:lhapdf} discusses
refinements to the \LHAPDF library, including both intrinsic performance improvements
and the importance of efficient call strategies.
Section~\ref{sec:improvs} details improvements of the \Sherpa event generator.
Section~\ref{sec:perfimprovs} quantifies the impact of our modifications.
In Sec.~\ref{sec:future} we discuss possible
future directions for further improvements of the two software packages,
and Sec.~\ref{sec:conclusion} provides an outlook.

\section{\texorpdfstring{\LHAPDF}{LHAPDF} performance bottlenecks and improvements}
\label{sec:lhapdf}

\newcommand{\pdf}[2][]{\ensuremath{f_{#2}^{#1}}}

While the core machinery of event generators for high-energy collider physics is framed in terms of partonic
scattering events, real-world relevance of course requires that the matrix elements be
evaluated for colliding beams of hadrons. This is typically
implemented through use of the collinear factorisation formula for the differential
cross section about a final-state phase-space configuration $\Phi$,
\begin{equation}
  \label{eq:factorization}
  \done{\sigma({h_1 h_2 \to n})}
  =
  \sum_{a,b}
  \, \int_{0}^{1} \! \done{x_a}
  \, \int_{0}^{1} \! \done{x_b}
  ~
  \pdf[h_1]{a}(x_a,\muF) \, \pdf[h_2]{b}(x_b,\muF)\,
  \done\hat\sigma_{ab\to n}(\Phi, \muR, \muF) \, ,
%
%
\end{equation}
where $x_{a,b}$ are the light-cone momentum fractions of the two incoming
partons $a$ and $b$ with respect to their parent hadrons $h_1$ and $h_2$,
and $\muRF$ are the renormalisation and factorisation scales,
respectively. Assuming negligible transverse motion of the partons, this formula
yields the hadron-level differential cross section $\done{\sigma}$ as an
integral over the initial-state phase-space, summed over $a$ and $b$,
weighting the differential squared matrix-element
$\mathrm{d}\hat\sigma$ by the collinear parton densities (PDFs) $f$ for the incoming beams.
These PDFs satisfy the evolution equations~\cite{Dokshitzer:1977sg,Gribov:1972ri,
  Lipatov:1974qm,Altarelli:1977zs}
\begin{equation}
  \label{eq:dglap}
  \frac{{\rm d}\ln f_{a}^h(x,t)}{{\rm d}\ln t}=
  \sum_{b=q,g}\int_0^1\frac{{\rm d} z}{z}\,\frac{\alpha_s}{2\pi}
  P_{ab}(z)\,\frac{f_{b}^h(x/z,t)}{f_{a}^h(x,t)}\;,  
\end{equation}
with the evolution kernels, $P_{ab}(z)$, given as a power
series in the strong coupling, $\alpha_s$.

In MC event-generation, the integrals in Eqs.~\eqref{eq:factorization}
and~\eqref{eq:dglap} are replaced by MC rejection sampling, meaning
that a set of PDF values $\pdf[h]{a}(x,\muF)$ must be evaluated at every
sampled phase-space point, for both beams.
PDFs are hence among the most intensely called functions within an
event generator code, comparable with the partonic matrix-element itself.
In particular, Eq.~\eqref{eq:dglap} is iteratively solved by the backward
evolution algorithm of initial-state parton showers~\cite{Sjostrand:1985xi},
requiring two PDF calls per trial emission~\cite{Ellis:1996mzs}.

This intrinsic computational load is exacerbated by the additional factors that
1)~the non-perturbative PDFs are not generally available as closed-form
expressions, but as discretised grids of $\pdf{}{}(x_i,Q^2_i)$ values obtained from
fits to data via QCD scale-evolution, and 2)~the PDF fits introduce many new
sources of systematic uncertainty, which are typically encoded via
$\mathcal{O}(10\text{--}100)$ alternative sets of PDF functions to be evaluated
at the same $(x,Q^2)$ points.  In LHC MC-event production, these grids are
interpolated to provide PDF values and consistent values of the running
coupling, $\alphaS$, through continuous $(x,Q^2)$ space by the \LHAPDF library.

The starting point for this work is \LHAPDF version 6.2.3, the C++ \LHAPDF6
lineage being a redevelopment of the Fortran-based \LHAPDF~$\le 5$ series. The
Fortran series relied on each PDF fit being supplied as a new subroutine by the
fitting group; in principle these used a common memory space across sets, but in
practice many separate such memory blocks were allocated, leading to
problematically high memory demands in MC-event production. The C++ series has a
more restrictive core scope, using dynamic memory allocation and a set of common
interpolation routines to evaluate PDF values from grids encoded in a standard
data format. Each \emph{member} of a collinear PDF \emph{set} is a set of
functions $f_a^h(x, Q^2)$ for each active parton flavour, $a$, and is
independently evaluated within \LHAPDF.

The most heavily used interpolation algorithm in \LHAPDF is a 2D local-cubic
polynomial~\cite{numrecipes} in $(\log x, \log Q^2)$ space, corresponding to
a composition of 1D cubic interpolations in first the $x$ and then the $Q^2$
direction on the grid. As each 1D interpolation requires the use of four
$\pdf{a}(x_i,Q_j^2)$ knot values, naively 16 knots are
needed as input to construct 4 values at the same $x$ value, used as the
arguments for the final 1D interpolation in $\log Q^2$. The end result is a
weighted combination of the PDF values on the 16 knots surrounding the
interpolation cell of interest, with the weights as functions of the position of
the evaluated point within the cell.

\subsection{PDF-grid caching}
The first effort to improve \LHAPDF's evaluation efficiency was motivated by the
sum over initial-state flavours in Eq.~\eqref{eq:factorization}, implying that
up to 11 calls (for each parton flavour, excluding the top quark) may be made
near-consecutively for a fixed $(x,Q^2)$ point within the same PDF.

If such repeated calls use the same $(x,Q^2)$ knot positions for all flavours
(which is nearly always the case), much of the weight computation described
above can be cached and re-used with a potential order-of-magnitude gain. Such a
caching was implemented, with a dictionary of cyclic caches stored specific to
each thread and keyed on a hash-code specific to the grid spacing: this ensures
that the caching works automatically across different flavours if they use the
same grid geometry but does not return incorrect results should that assumption
be incorrect. This implementation also has the promising side-effect that,
if the set of fit-variation PDFs also use the same grid spacing as the nominal PDF,
consecutive accesses of the same $(x,Q^2)$ across possibly hundreds of PDFs
would also automatically benefit from the caching.

\begin{figure}[t!]
  \centering
  \includegraphics[width=0.7\textwidth]{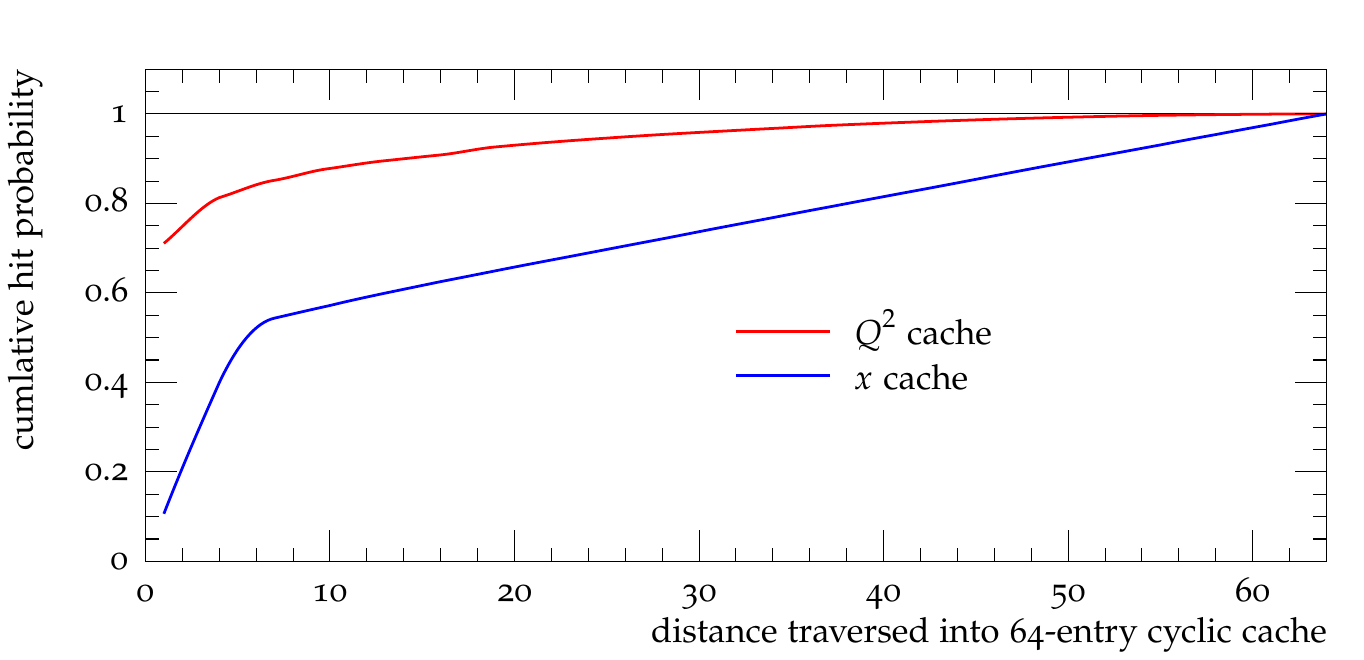}
  \caption{
    Cumulative probability of obtaining cache-hit as a function of search depth into a
    64-entry cyclic cache for calls to $x$ and $Q^2$ by \Sherpa when generating
    $e^+ e^-$+jet MC events. As a proportion of all calls which resulted in a cache-hit.
    \label{fig:lhapdf:wjet:cachehits}
  }
\end{figure}

The practicality of a cache implementation in \LHAPDF (with no restructuring of
the call patterns from \Sherpa) was investigated using the
$e^+ e^-$+jets setup described below and a 64-entry cyclic cache.
This cache is too large to obtain
any performance benefits but was useful to explore the caching behaviour.
57\% of $x$ and 54\% of $Q^2$ lookups were located within the 64-entry cache.
Of these successful cache-hits, the cumulative probability of an $x$ hit rose
linearly from 10\% in the first check to 50\% by the 6th check before slowing
down (90\% by the 51st check), as illustrated in
Figure~\ref{fig:lhapdf:wjet:cachehits}.  For $Q^2$, the cumulative probability
was already at 80\% by the third check (90\% by the 13th check).

Despite this promise, this caching feature as implemented in \LHAPDF~6.3.0
transpired to add little if anything in practical applications with \Sherpa
generation of these ATLAS-like $e^+ e^-$+jets MC events. With a cache depth of 4, the time
spent in \LHAPDF in the call-stack reduced marginally by a relative 5\%, this overall reduction is small due to 
29\% of the time spent in \LHAPDF now being under the newly added \texttt{\_getCacheX} and 
\texttt{\_getCacheQ2} functions. 
This indicates that, given the \Sherpa request pattern, the cost of executing
the caching implementation is somewhat comparable to the cost of re-interpolating the quantity.

This experience of caching as a strategy to reduce PDF-interpolation overheads
in realistic LHC use-cases highlights the importance of well-matched PDF-call
strategies in the event generator. We return to this point later.

\subsection{Memory structuring and return to multi-flavour caching}
The C++ rewrite of \LHAPDF placed emphasis on flexibility and ``pluggability'' of
interpolators to accommodate fitting groups' requirements, allowing the use of
non-uniform grid spacings, functional discontinuities across flavour thresholds,
and even different grids for each parton flavour~\cite{Buckley:2014ana},
at the cost of a fragmented memory layout.
However, much of this flexibility has in practice gone unused.

By disabling the possibility to have fragmented knots for differing flavours,
the knots are now stored in a single structure for all flavours.
Similarly, the PDF grids are stored in a combined data-structure. This will allow
for very efficient caching and even memory accesses due to the contiguous
memory layout.

With the observed shortcomings in the caching-strategy implemented in \LHAPDF~6.3.0, as described above,
in \LHAPDF~6.4.0, the caching mechanism focuses on multi-flavour PDFs that are called for explicitly.
In this case, large parts of the computations
can be shared between the different flavour PDF (for example finding the right knot-indices
and computing spacings) due to the fact that the grids have been
unified.
In principle, the caching of shared computations among the variations is still desirable,
given that many variations share grids. However as discussed above, the call strategy of the generator
then has to be structured (or, restructured) with this in mind in order to make this caching efficient.

\subsection{Finite-difference precomputations}
Additionally to the reworked caching strategy, \LHAPDF~6.4.0 pre-computes parts of the computations
and stores the results.
Due to the way the local-cubic polynomial interpolation is set up, the first set of interpolations
are always computed along the grid lines. Since these are always the same, in \LHAPDF~6.4.0
the coefficients of the interpolation polynomial are pre-computed for the grid-aligned interpolations.
This comes with the drawback of the additional memory space that is required to store the coefficients,
but it also reduces the interpolation to simply the evaluation of a cubic polynomial
(compared to first constructing said polynomial, and then evaluating it).
The precomputations reduce the number of ``proper'' interpolations
(in the sense that the interpolation polynomial has to be constructed) from five to one.

Because of these precomputations and the above described memory restructurings,
computing the PDF becomes up to a factor of $\sim\!3$ faster for a single flavour,
and with the combination of the multi-flavour caching, computing the PDFs for for
all flavours becomes roughly $\sim\!10$ faster.

\section{\texorpdfstring{\Sherpa}{Sherpa} performance bottlenecks and improvements}
\label{sec:improvs}

The computing performance of various LHC event generators was investigated
in a recent study performed by the HEP software foundation~\cite{
  HSFPhysicsEventGeneratorWG:2020gxw,HSFPhysicsEventGeneratorWG:2021xti,
  HEPSoftwareFoundation:2020daq}. This comparison prompted a closer inspection
of the algorithms used and choices made in the Sherpa program.
In this section we will briefly review the computationally most demanding
parts of the simulation, provide some background information on the physics
models, and offer strategies to reduce their computational complexity.

We will focus on the highly relevant processes $pp\to\ell^+\ell^-+\text{jets}$
and $pp\to t\bar{t}+\text{jets}$, described in detail in Sec.\
\ref{sec:perfimprovs}. They are typically simulated using NLO
multi-jet merged calculations with EW virtual corrections and include scale
as well as PDF variations. The baseline for our simulations is the \Sherpa
event generator, version~2.2.11~\cite{Bothmann:2019yzt}. In the typical
configuration used by the ATLAS experiment, it employs the \Comix matrix element
calculator~\cite{Gleisberg:2008fv} to compute leading-order cross sections
with up to five final-state jets in $pp\to\ell^+\ell^-+\text{jets}$ and
four jets in $pp\to t\bar{t}+\text{jets}$. Next-to-leading order
precision in QCD is provided for up to two jets in $pp\to\ell^+\ell^-+\text{jets}$
and up to one jet in $pp\to t\bar{t}+\text{jets}$ with the help of
the \OpenLoops library~\cite{Cascioli:2011va,Buccioni:2019sur} for virtual
corrections and an implementation of Catani-Seymour dipole subtraction
in \Amegic~\cite{Gleisberg:2007md} and \Comix.
The matching to a Catani-Seymour based parton shower~\cite{Schumann:2007mg}
is performed using the \SMCatNLO technique~\cite{Hoeche:2011fd,Hoeche:2012fm},
an extension of the \MCatNLO matching method~\cite{Frixione:2002ik} that implements colour
and spin correlations in the first parton-shower emission, in order to
reproduce the exact singularity structure of the hard matrix element.
In addition, EW corrections and scale-, $\alphaS$- and PDF-variation
multiweights are implemented
using the techniques outlined in~\cite{Kallweit:2015dum,Brauer:2020kfv,
  Bothmann:2016nao}.
A typical setup includes of the order of two hundred multiweights,
most of which correspond to PDF variations.

\begin{figure}[tb!]
  \includegraphics[width=\textwidth,clip,trim=0 17.5mm 0 0]{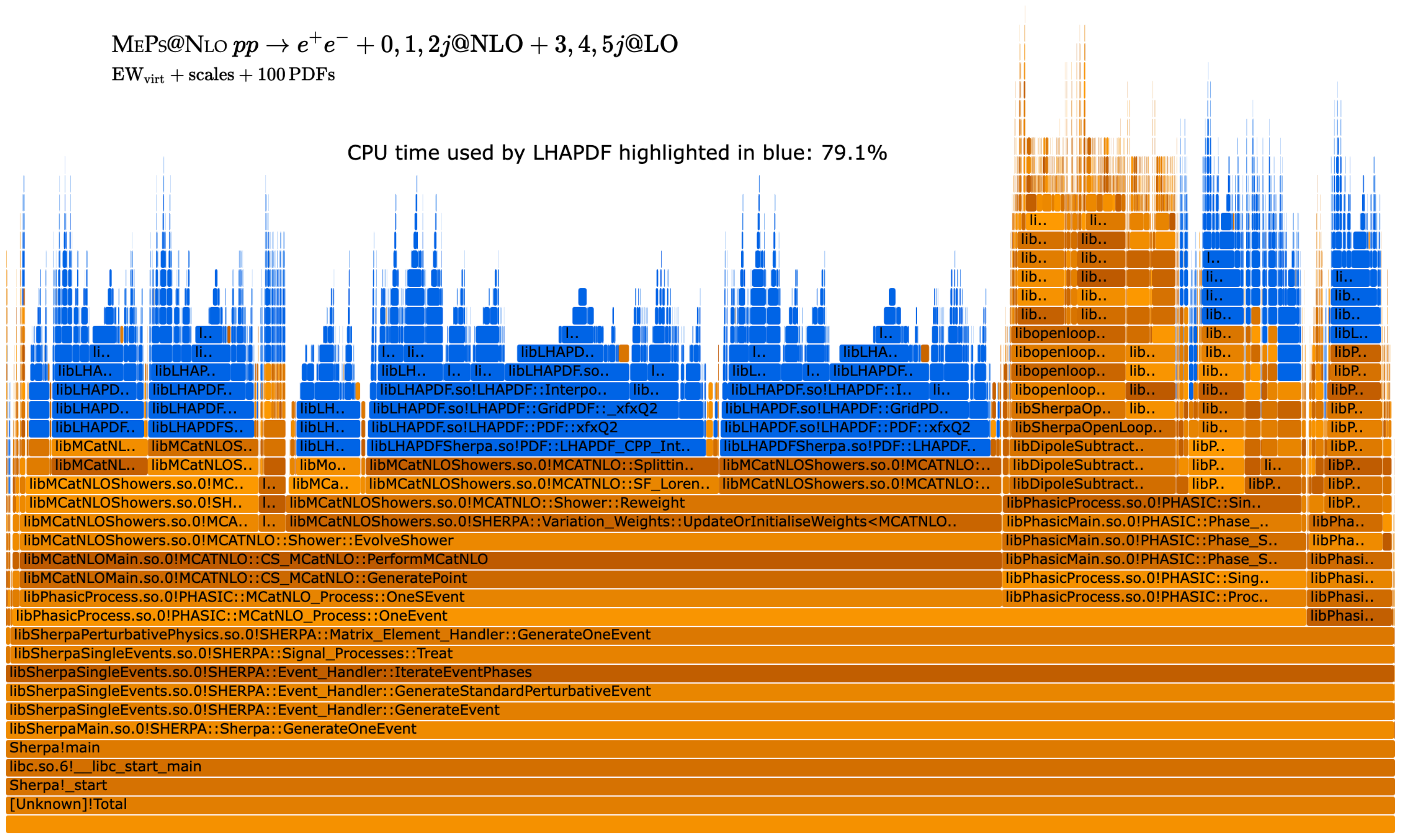}
  \caption{
    CPU profile of 1000 MC unweighted $pp\to e^ + e^-$+jet events generated by \Sherpa 2.2.11
    interfaced with \LHAPDF 6.2.3. The 79\% of run-time spent within \LHAPDF
    in the call-stack is highlighted in blue.
    \label{fig:lhapdf:wjet:flamegraph}
  }
\end{figure}
We visualize the imperfect interplay between \Sherpa and \LHAPDF
in Fig.~\ref{fig:lhapdf:wjet:flamegraph}. For this test, \Sherpa~2.2.11
was compiled against \LHAPDF~6.2.3 and
\OpenLoops~2.1.2~\cite{Cascioli:2011va,Buccioni:2019sur}.
The performance of generating 1000 partially unweighted MC events was then
profiled with the Intel$^\text{\textregistered}$\xspace VTune\texttrademark\xspace
profiler running on a single core of a \SI{2.20}{GHz} Intel$^\text{\textregistered}$ 
Xeon$^\text{\textregistered}$ E5-2430.
The \Sherpa run card contains a representative
$pp \to e^+ e^- + 0,1,2j$@NLO$+3,4,5j$@LO setup at $\sqrt{s} = 13$~\TeV, including
electroweak virtual corrections as well as reweightings to different PDFs and scales;
comparable to the setup used in production by the ATLAS collaboration at the time.
The total processing time was around 18.5 hours.

The obtained execution profile is visualized in
Figure~\ref{fig:lhapdf:wjet:flamegraph} as a flame-graph~\cite{flamegraph} where
the proportion of the $x$-axis reflects the proportion of wall-time spent inside
a given function, and where the call-stack extends up the $y$-axis. Calls from
\Sherpa into the \LHAPDF library are highlighted in blue. In total, \SI{79}{\percent}
of the execution time was spent in \LHAPDF, with
\texttt{libLHAPDFSherpa.so!PDF::LHAPDF\_CPP\_Interface::GetXPDF} representing
the dominant interface call.

In the following, we discuss in detail the major efficiency improvements
that have been implemented on the \Sherpa side,
including the solution to spending so much execution time within \LHAPDF.
In addition to the major changes, also some minor improvements have been developed,
which account for a collective runtime savings of \SIrange{5}{10}{\percent}.
A notable example is the introduction of a cache
for the partonic channel selection weights,
reducing the necessity to resolve virtual functions in inheritance structures.

\subsection{Leading-colour matched emission}
\label{sec:improvs:lcmatching}
A simple strategy to improve the performance of the \SMCatNLO
matching was recently discussed in~\cite{Danziger:2021xvr}.
Within the \SMCatNLO technique, one requires the parton shower to
reconstruct the exact soft radiation pattern obtained in the NLO result.
In processes with more than two coloured particles, this leads
to non-trivial radiator functions, which are given in terms of
eikonals obtained from quasi-classical currents~\cite{Bassetto:1984ik}.
Due to the involved colour structure of the related colour insertion
operators, the radiation pattern can typically not be captured
by standard parton shower algorithms. The \SMCatNLO technique relies
on weighted parton showers~\cite{Hoeche:2009xc} to solve this problem.
As both the sign and the magnitude of the colour correlators can differ
from the Casimir operator used in leading colour parton showers,
the weights can become negative and are in general
prone to large fluctuations that need to be included in the overall
event weight, thus lowering the unweighting efficiency
and reducing the statistical power of the event sample.

This problem can be circumvented
by assuming that experimentally relevant observables will likely
not be capable of resolving the details of soft radiation, and that
colour factors in the collinear (and soft-collinear) limit are given
in terms of Casimir operators. This idea is also used in the
original \MCatNLO method~\cite{Frixione:2002ik} to enable the matching
to parton showers which do not have the correct soft radiation pattern.
Within \Sherpa, the \SMCatNLO matching is simplified to an \MCatNLO
matching, dubbed \LCMCatNLO here, using the setting \NLOCSSPSMODEi{1}.
Without further colour correlators, no additional weight is
added, making the unweighting procedure more efficient.

With \SMCatNLO, the
parton shower needs information about soft-gluon insertions into the
Born matrix element, which makes the first step of the parton shower
dependent on the matrix-element generator. In fact, within Sherpa
the first emission is generated as part of the matrix-element simulation
by default. When run in \LCMCatNLO mode, the dependence of the parton shower
on the matrix-element generator does not exist. Using the flag \NLOCSSPSMODEi{2},
the user can then include the generation of the first emission
into \Sherpa's standard Catani-Seymour shower (\CSS).
We will call this configuration \LCMCatNLOCSS in the following.
The first emission is then performed after the unweighting step,
such that it is not generated any longer for events that might
eventually be rejected. This simplification leads to an additional speedup.

The above argument is also employed for spin correlations in
collinear gluon splittings, which are normally included in
\SMCatNLO. Assuming experimentally relevant observables to be
insensitive to it, we reduce the corresponding spin-correlation
insertion operators to their spin-averaged counterparts present
in standard parton shower algorithms in the \LCMCatNLO and
\LCMCatNLOCSS implementations.

\subsection{Pilot-run strategy}
\label{sec:improvs:pilotrun}

In the current implementation of \Sherpa's
physics modules and interfaces to external libraries,
physical quantities and coefficients that are needed
later in the specified setup, \eg to calculate
scale and PDF variations and other alternative
event weights, are calculated
when the program flow passes through the specific
module or interface.
While this is the most efficient strategy for weighted
event generation and allows for easy maintainability
of the implementation, it is highly inefficient in
unweighted event generation and in fact responsible for most
of the large fraction of computing time spent in \LHAPDF calls
in Fig.~\ref{fig:lhapdf:wjet:flamegraph}.
This is because the unweighting is based solely on the
nominal event weight and these additional quantities and
coefficients will only be used once an event has been
accepted and are thus calculated needlessly for
events that are ultimately rejected in the unweighting
step.

To improve code performance for unweighted event
generation, we thus introduce a pilot run.
We reduce the number of coefficients to be calculated
to a minimal set until an event has been accepted.
Once such an event is found, we recompute this
exact phase space point including all later-on
desired coefficients.
Thus, the complete set of variations and alternative
event weights is computed only for the accepted event,
while no unnecessary calculations are
performed for the vast number of ultimately rejected
events.

The pilot-run strategy is introduced in \Sherpa-2.2.12
and is used automatically for (partially) unweighted
event generation that includes variations.

\subsection{Analytic virtual corrections}
\label{sec:improvs:analytics}
Over the past decades fully numerical techniques have been
developed to compute nearly arbitrary one-loop amplitudes~\cite{
  Berger:2008sj,Berger:2009ep,Berger:2010vm,Berger:2010zx,Ita:2011wn,
  Bern:2013gka,Hirschi:2011pa,Alwall:2014hca,Cascioli:2011va,Buccioni:2017yxi,
  Buccioni:2019sur,Badger:2010nx,Badger:2012pg,Cullen:2011ac,Cullen:2014yla,
  Actis:2012qn,Actis:2016mpe,Denner:2017vms,Denner:2017wsf}.
The algorithmic appeal of these approaches
makes them prime candidates for usage in LHC event generators.
Their generality does, however, come at the cost of reduced
computing efficiency in comparison to known analytic results.
In addition, the numerical stability of automated calculations
can pose a problem in regions of phase space where partons
become soft and/or collinear, or in regions affected by thresholds.
Within automated approaches, these numerical instabilities can
often only be alleviated by switching to higher numerical precision,
while for analytic calculations, dedicated simplifications or series
expansions of critical terms can be performed.
For the small set of standard candle processes at the LHC that require
high fidelity event simulation, one may therefore benefit immensely
from the usage of the known analytic one-loop amplitudes.

Most of the known analytic results of relevance to LHC physics
are implemented in the Monte Carlo for FeMtobarn processes
(\MCFM)~\cite{Campbell:1999ah,Campbell:2011bn,Campbell:2015qma,Campbell:2019dru}.
A recent project made these results accessible for event generation
in standard LHC event generators~\cite{Campbell:2021vlt} through a
generic interface based on the Binoth Les Houches
Accord~\cite{Binoth:2010xt,Alioli:2013nda}. A similar interface
to analytic matrix elements was provided in the \BlackHat
library~\cite{Berger:2008sj}.

Since \MCFM does not provide the electroweak one-loop corrections
which are relevant for LHC phenomenology in the high transverse
momentum region, we use the interface to analytic matrix elements
primarily for the pilot runs before unweighting. The full calculation,
including electroweak corrections, is then performed with the help of
\OpenLoops. This switch is achieved by the setting \UnwgtGenX{MCFM}.

\subsection{Extending the pilot run strategy to reduce jet clustering}
\label{sec:improvs:pilotscale}
For multijet-merged runs using the
\CKKW-L algorithm~\cite{Catani:2001cc,Lonnblad:2001iq},
the final-state configurations are re-in\-ter\-pre\-ted
as having originated from a parton cascade~\cite{Andre:1997vh}.
This is called clustering, and the resulting
parton shower history is used to choose
an appropriate renormalisation scale
for each strong coupling evaluation
in the cascade,
thus resumming higher-order corrections to soft-gluon radiation~\cite{Amati:1980ch}.
This procedure is called $\alpha_s$-reweighting.
The clustering typically requires the determination
of all possible parton-shower histories,
to select one according to their relative probabilities~\cite{Andre:1997vh,Lonnblad:2001iq}.
The computational complexity therefore grows quickly
with the number of final-state particles~\cite{Hoche:2019flt}.
It can take a significant share of the computing time
of a multi-jet merged event,
as we will see in Sec.~\ref{sec:perfimprovs}.

To alleviate these problems, we have implemented a scale setting
which uses a surrogate scale choice for the pilot events,
while the $\alpha_s$ reweighting is only done
once an event has been accepted,
thus avoiding the need to determine clusterings
for the majority of trial events.
Contrary to the improvements discussed in Sec.~\ref{sec:improvs:pilotrun},
this changes the weight of the event.
To account for this change,
the ratio of the two different cross sections
before and after the unweighting
must either be used as an additional event weight,
or as the basis of an additional second unweighting procedure.
In our implementation, we chose the former procedure,
expecting a rather peaked weight distribution,
such that additional event processing steps
(such as a detector simulation)
retain a high efficiency
even though the events do not carry a constant weight.

\section{Observed performance improvements}
\label{sec:perfimprovs}

In this section we investigate the impact of the
performance improvements detailed in
Secs.\ \ref{sec:lhapdf} and \ref{sec:improvs}.
As test cases we use the following setups:
\begin{description}
  \item[\eeproc] \ \\
            Drell-Yan production at 13\,TeV at the LHC.
            We bias the unweighted event
            distribution in the maximum of the scalar sum of 
            all partonic jet transverse momenta ($H_\mathrm{T}$) 
            and the transverse momentum of the lepton pair ($p_\mathrm{T}^V$),
            leading to a statistical over-representation of multijet events.
  \item[\ttproc] \ \\
            Top-pair production at 13\,TeV at the LHC.
            We bias the unweighted event
            distribution in the maximum of the scalar sum of 
            all non-top partonic jet transverse momenta ($H_\mathrm{T}$) 
            and the average top-quark ($(p_\mathrm{T}^t + p_\mathrm{T}^{\bar{t}})/2$),
            leading to a statistical over-representation of multijet events.
\end{description}
In each case, the different multiplicities at leading and
next-to-leading order are merged using the \MEPSatNLO algorithm detailed
in \cite{Hoeche:2012yf,Gehrmann:2012yg,Hoeche:2014rya}.
The setups for both processes
reflect the current usage of \Sherpa in the ATLAS experiment,
and have also been used for a study on the reduction
of negative event weights~\cite{Danziger:2021xvr}.
The corresponding runcards can be found in App.\ \ref{app:runcards}.

The performance is measured in five variations of the two process setups,
with an increasing number of additionally calculated event weights
corresponding to QCD variations (scale factors and PDFs)
and approximative EW corrections (\EWvirt):
\begin{description}
  \item[no variations] \ \\
    No variations, only the nominal event weight is calculated.
  \item[\EWvirt]\ \\
    Additionally, \EWvirt\ corrections are calculated.
    This requires the evaluation of the EW virtual correction
    and subleading Born corrections.
    In particular the evaluation of the virtual part
    has a significant computational cost.
  \item[\EWvirt+scales]\ \\
    Additionally, 7-point scale variations are evaluated,
    both for the matrix-element and the parton-shower parts
    of the event generation~\cite{Bothmann:2016nao}.
    This includes the re-evaluation of
    couplings (when varying the renormalisation scale)
    and PDFs (when varying the factorisation scale),
    of which the latter are particularly costly.
  \item[\EWvirt+scales+100 PDFs]\ \\
    Additionally, variations
    are calculated for 100 Monte-Carlo replica of the used
    PDF set (\texttt{NNPDF30\_nnlo\_ as\_0118}~\cite{NNPDF:2014otw}).
    This again requires the re-evaluation of the PDFs
    both in the matrix element and the parton shower.
    As for the scale variations, the cost scales approximately linearly
    with the number of variations.
    Note that this setup variation is closest to
    what would be typically used in an ATLAS
    vector-boson or top-pair productions setup,
    which might however feature a number of PDF variations
    which is closer to 200.
  \item[\EWvirt+scales+1000 PDFs]\ \\
    This setup variation is similar to the previous,
    with the only difference being that
    the 1000 instead of the 100 Monte-Carlo replica error set
    of the \texttt{NNPDF30\_nnlo\_as\_0118} PDF set is used.
\end{description}

The impact of the performance improvements is investigated in seven steps,
with each step adding a new improvement as follows:
\begin{description}
  \item [\MEPSatNLO baseline] \ \\
      This is our baseline setup, using the pre-improvement
      versions of \Sherpa 2.2.11 and \LHAPDF 6.2.3, \ie
      using the CKKW scale setting procedure throughout
      as well as the standard \SMCatNLO matching technique.
      All one-loop corrections are provided by \OpenLoops.
  \item [\raisebox{2pt}{$\drsh$} \LHAPDF 6.4.0] \ \\
      The version of \LHAPDF is increased to \LHAPDF 6.4.0, implementing
      the improvements of Sec.\ \ref{sec:lhapdf}.
  \item [\raisebox{2pt}{$\drsh$} \LCMCatNLO] \ \\
      The full-colour spin-correlated \SMCatNLO algorithm is reduced
      to its leading-colour spin-averaged
      cousin, \LCMCatNLO, which however is still applied before the unweighting.
      Note that this is the only step where a physics simplification occurs.
      For details \see Sec.\ \ref{sec:improvs:lcmatching}.
  \item [\raisebox{2pt}{$\drsh$} pilot run] \ \\
      The pilot run strategy of Sec.\ \ref{sec:improvs:pilotrun} is enabled,
      minimising the number of coefficients and variations
      needlessly computed for events that are going to be rejected
      in the unweighting step.
  \item [\raisebox{2pt}{$\drsh$} \LCMCatNLOCSS] \ \\
      The \LCMCatNLO matching is moved into the standard \CSS parton shower,
      \ie it is now applied after the unweighting.
  \item [\raisebox{2pt}{$\drsh$} \MCFM] \ \\
      During the pilot run,
      the automatically generated one-loop QCD matrix elements provided
      by \OpenLoops are replaced by the manually highly
      optimised analytic expressions encoded in \MCFM.
      Once the event is accepted, \OpenLoops continues to provide all
      one-loop QCD and EW corrections, \see Sec.\ \ref{sec:improvs:analytics}.
  \item [\raisebox{2pt}{$\drsh$} pilot scale]
      Events are unweighted using a simple scale that depends solely on
      the kinematics of the final state and, thus, does not require
      a clustering procedure. The correct dependence on the actual
      factorisation and renormalisation scales determined through
      the CKKW algorithm is then restored through a residual event weight.
      For details \see Sec.\ \ref{sec:improvs:pilotscale}.
\end{description}

For the benchmarking, a dedicated computer is used with no additional computing load present 
during the performance tests.
The machine uses an Intel$^\text{\textregistered}$ Xeon$^\text{\textregistered}$ E5-2430
with a 2.20 GHz clock speed.
Local storage is provided through a RAID~0 array
of a pair of Seagate$^\text{\textregistered}$ $2.5^{\prime\prime}$ \SI{600}{GB} \SI{10}{kRPM} hard-drive
with a \SI{12}{Gb/s} SAS interface.
Six \SI{8}{GB} DD3 dual in-line memory modules with
1333 million transfers per second are used for dynamic volatile memory.

\begin{figure}[t!]
  \includegraphics[width=\textwidth]{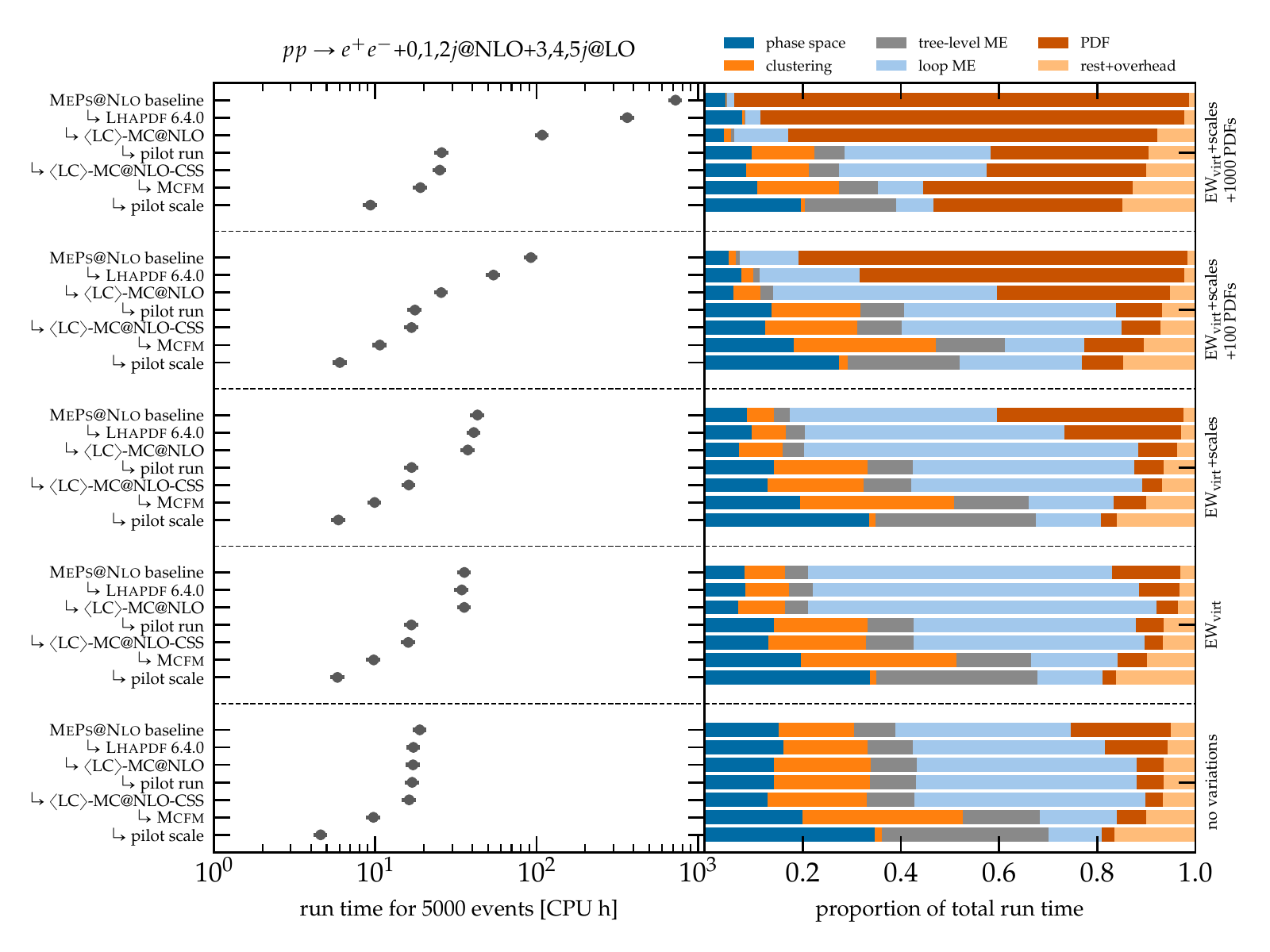}
  \caption{
    Reduction in overall run time for different performance improvements,
    combined with the breakdown of the overall run time into a high-level
    calculation composition.
    The timing is assessed by producing 5000 particle-level events for
    $pp \to e^+e^- + 0,1,2j$@NLO${}+3,4,5j$@LO using \MEPSatNLO.
    The scaling with the number of additional variation weights is
    benchmarked through a few representative setup configurations.
    \label{fig:perfimprovs:zjet:combined}
  }
\end{figure}



\paragraph*{\eeproc.}
We begin our analysis by examining the behaviour of the $e^+e^- + \text{jets}$ setup.
Figure~\ref{fig:perfimprovs:zjet:combined}
shows the impact of each improvement on the total run time to
generate 5000 unweighted events on the left side,
and the composition of these run times
for each of the seven steps, respectively, on the right side.
For the total run times,
horizontal error bars indicate a
\SI{10}{\percent} uncertainty estimate.

First, we note that using \LHAPDF 6.4
reduces the overall run time by about \SIrange{40}{50}{\percent}
when many PDF variations are used,
i.e.\ for the setup variants with 100 and 1000 PDF variations.
Unsurprisingly, the proportion of total runtime
dedicated to PDF evaluation shrinks accordingly.

The effect of additionally enabling \LCMCatNLO
scales with the number of PDF and scale variations,
which also determines the number
of required \MCatNLO one-step shower variations required.
Hence, for \EWvirt+scales, it gives a speed-up of about \SI{10}{\percent},
while for the setup with 1000 PDF variations, more than a factor of three is gained.

The biggest impact (apart from the ``no variations'' setup variant)
is achieved when also enabling the pilot run.
It removes the overhead of calculating variations nearly entirely,
such that the resulting runtimes are then very comparable
across all setup variants.
Only when calculating 1000 PDF variations there is still a sizeable
increase of about \SI{40}{\percent} in runtime,
compared to the ``no variations'' variant.

Additionally moving the matched first shower emission into the normal
\CSS shower simulation, \LCMCatNLOCSS,
gives a speed-up of \SIrange{5}{10}{\percent} for all setup variants.

Then, switching to use \MCFM for pre-unweighting loop calculations
gives another sizeable reduction in runtime by about \SI{80}{\percent}.
This reduction is only diluted somewhat in the 1000 PDF variation case,
given the sizeable amount of time that is still dedicated to calculating variations
of the unweighted events.

Lastly, we observe another \SIrange{50}{60}{\percent} reduction
of the required CPU time when choosing a scale definition that does
not need to reconstruct the parton shower history to determine the
factorisation and renormalisation scales of a candidate event in the
pilot run.\footnote{\label{footnote:loop_optimisation}%
  From the runtime composition on the right side of Fig.~\ref{fig:perfimprovs:zjet:combined},
  one can see that this is not entirely due to the minimised time spent in the clustering,
  but also due to a somewhat reduced time usage in the loop matrix elements.
  This stems from an improved unweighting efficiency
  for Born-like configuration including virtual corrections
  when optimising the event generation using the simplified pilot scale,
  which is likely due to the increased stability of the pilot scale.}
It has to be noted though that the correction to the proper CKKW
factorisation and renormalisation scales induces a residual weight,
\ie a broader weight distribution, leading to a reduced statistical
power of the resulting sample of the same number of events.
We will discuss this further below.

The overall reduction in runtime for the setup variants is
summed up in Tab.~\ref{tab:runtime}.
\begin{table}
  \begin{center}
    \begin{tabular}{lcrrrcrrr}\toprule
    & \hspace{2\tabcolsep} 
    & \multicolumn{3}{c}{$pp \to e^+e^-$ + jets}
    & \hspace{2\tabcolsep} 
    & \multicolumn{3}{c}{$pp \to t\bar{t}$ + jets} \\
    \cmidrule(r){3-5}
    \cmidrule(l){7-9}
    \addlinespace
    &
    & \multicolumn{3}{c}{runtime [CPU h/5k events]}  
    &
    & \multicolumn{3}{c}{runtime [CPU h/1k events]} \\
      setup variant & & old & new & speed-up &
    & old & new & speed-up \\ \midrule
      no variations            & &  20\,h & 5\,h &  4$\times$ & &  15\,h & 8\,h &  2$\times$ \\
      \EWvirt                  & &  35\,h & 5\,h &  6$\times$ & &  20\,h & 8\,h &  2$\times$ \\
      \EWvirt+scales           & &  45\,h & 5\,h &  7$\times$ & &  25\,h & 8\,h &  4$\times$ \\
      \EWvirt+scales+100 PDFs  & &  90\,h & 5\,h & 15$\times$ & &  55\,h & 8\,h &  7$\times$ \\
      \EWvirt+scales+1000 PDFs & & 725\,h & 8\,h & 78$\times$ & & 440\,h & 9\,h & 51$\times$ \\\bottomrule
  \end{tabular}
  \caption{\label{tab:runtime}%
    Overall reduction in run time for all performance improvements combined.
    The timing is assessed by producing 5000 particle-level events for
    $pp \to e^+e^- + 0,1,2j$@NLO${}+3,4,5j$@LO
    and 1000 particle-level events for
    $pp \to t\bar{t} + 0,1j$@NLO${}+2,3,4j$@LO,
    both at \MEPSatNLO.
    The scaling with the number of additional variation weights is benchmarked through a few representative setup configurations.}
  \end{center}
\end{table}
It is interesting to note that after applying all of the performance improvements,
there is no longer a single overwhelmingly computationally intense component left
in the composition shown in Fig.~\ref{fig:perfimprovs:zjet:combined}
(see the bottom line in each setup variant block):
None of the components in the breakdown use more than \SI{40}{\percent} of the runtime.
With the exception of the 1000 PDF variation setup variant,
the phase-space and tree-level ME components alone now
require more than \SI{50}{\percent} of the total runtime, such
that they need to be targeted for further performance improvements.
Also the virtual matrix elements (``loop ME'') are still sizeable
(approximately \SIrange{5}{10}{\percent} of the runtime),
albeit much smaller than the time spent on the remainder of the event generation.
However, from the perspective of the \Sherpa framework this is now irreducible
as the runtime is spent in highly optimised external loop matrix-element libraries,
and only when it is absolutely necessary.

\begin{figure}[t!]
  \includegraphics[width=\textwidth]{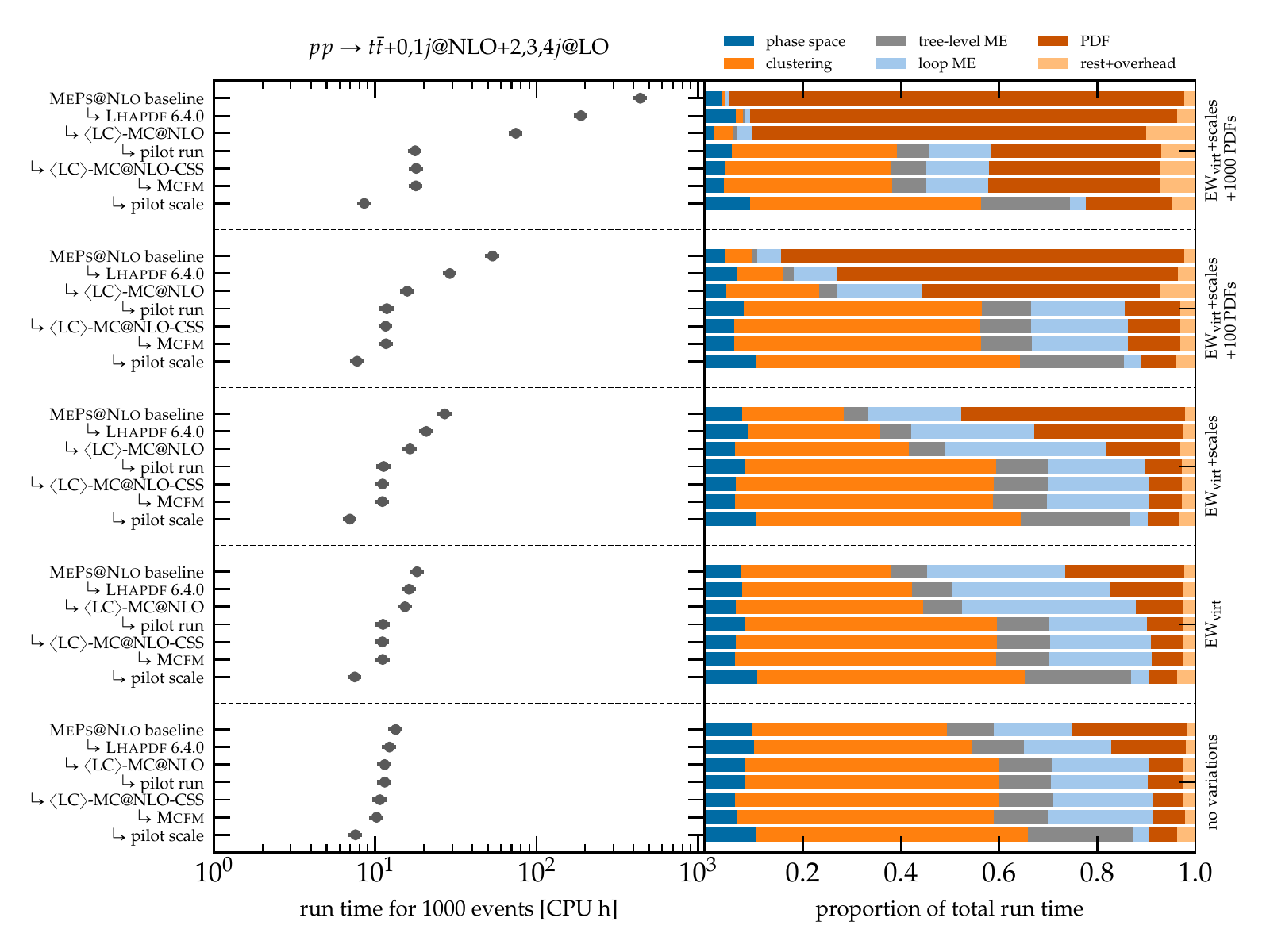}
  \caption{
    Reduction in overall run time for different performance improvements,
    combined with the breakdown of the overall run time into a high-level
    calculation composition.
    The timing is assessed by producing 1000 particle-level events for
    $pp \to t\bar{t} + 0,1j$@NLO${}+2,3,4j$@LO using \MEPSatNLO.
    The scaling with the number of additional variation weights is
    benchmarked through a few representative setup configurations.
    \label{fig:perfimprovs:ttjet:combined}
  }
\end{figure}

\paragraph*{\ttproc}
Following the analysis of the $e^+e^- + \text{jets}$ case,
we now present the breakdown of $t\bar t + \text{jets}$ run times and
their compositions
in Fig.\ \ref{fig:perfimprovs:ttjet:combined}.
Overall, the results are very similar.
The most striking difference in the runtime decomposition
is that the clustering part is about twice as large
compared to the $e^+e^- + \text{jets}$ case.
This is mainly related to the usage of a clustering-based
scale definition in the $\mathbb{H}$-events,
and also to the different structure of the core process.
In the $t\bar t$ case, the  initial state
is dominated by gluons instead of quarks,
and the core process comprises four partons instead of two.
Therefore, there are considerably more ways to cluster
a given jet configuration back into the core process.
Secondly, we find that the loop matrix elements
have a smaller relative footprint in the $t\bar t$ case,
which is due to them only being calculated to NLO accuracy
for up to one additional jet (as opposed to two additional jets
in the $e^+e^-$ case).

The speed-ups by the performance improvements are similar,
but the larger proportion of the clustering
and the smaller proportion of the loop matrix elements
results in the pilot run improvement and the
analytic loop matrix element improvement
having a smaller impact than in the $e^+e^-$ case.
Using the pilot scale also has a smaller effect than in to the $e^+e^-$ case: 
the simulation of $\mathbb{H}$-events requires the clustering to 
determine the parton shower starting scale as a phase space boundary
for their shower subtraction terms~\cite{Hoeche:2012yf}.
The large clustering component can only be removed 
if the $\mathbb{H}$-events are calculated using 
a dedicated clustering-independent scale definition, 
as is the case in the $e^+e^-$ setup.
Overall, the final runtime improvements
as reported in Tab.~\ref{tab:runtime}
are smaller than the ones for the $e^+e^-$ process,
but still very sizeable.

The most notable deviation in the improvement pattern
comes from switching to \MCFM for the unweighting step,
which only has a minor impact in the $t\bar{t}$ case.
This is due to the fact that only the $t\bar{t}$ process
is implemented in this library while the $t\bar{t}j$ process,
which is much more costly, has to be taken from \OpenLoops
throughout.

\paragraph*{Weight distribution for pilot scale}
\begin{figure}[tp]
  \includegraphics[width=0.49\textwidth]{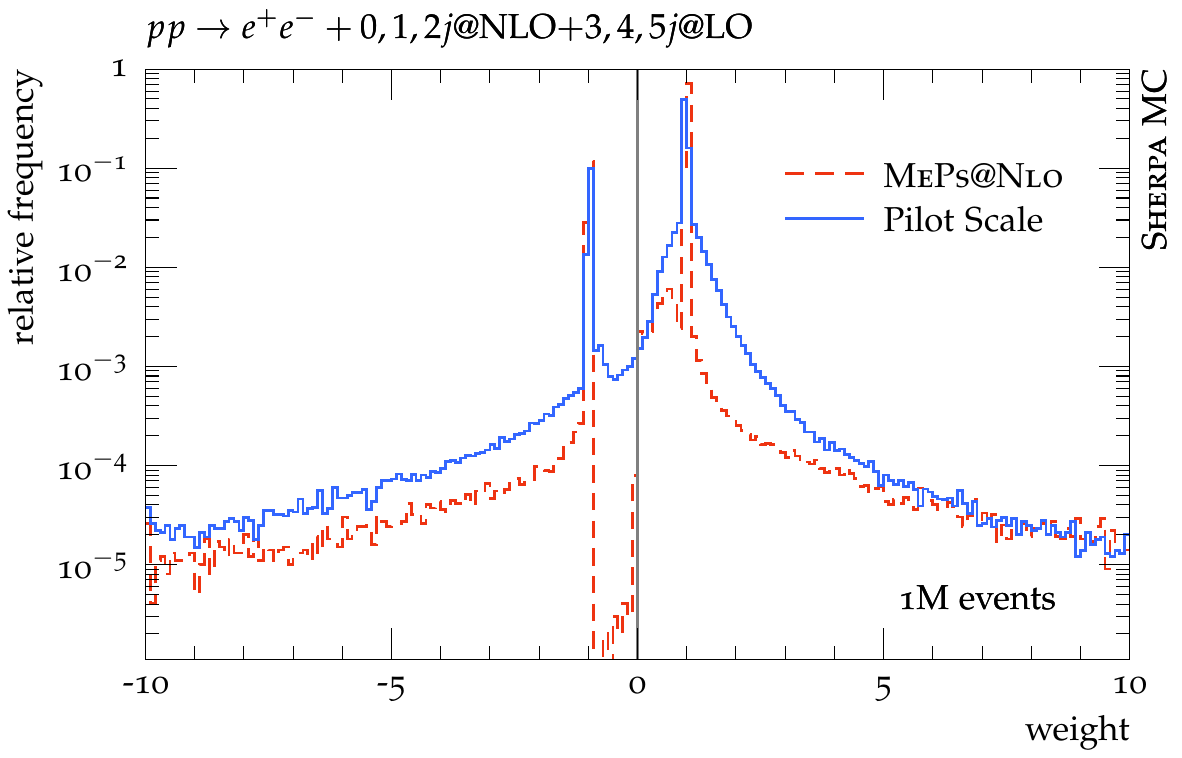}\hfill
  \includegraphics[width=0.49\textwidth]{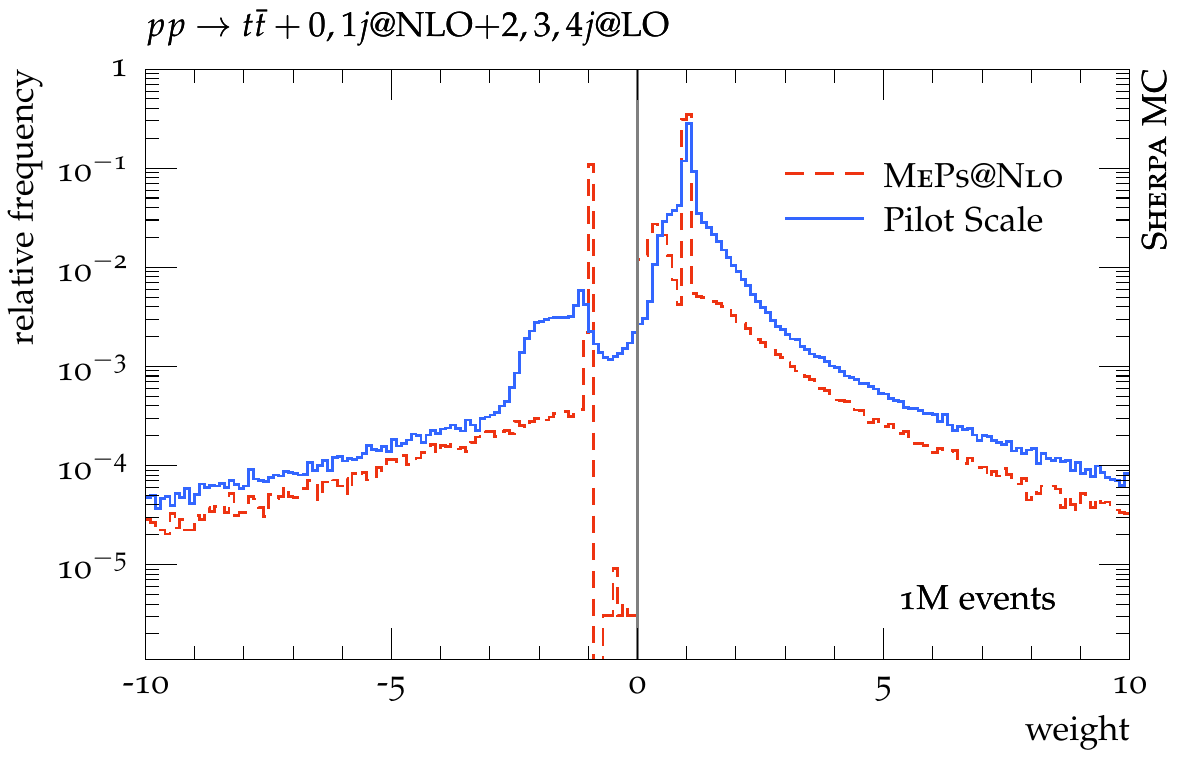}
  \caption{\label{fig:pilot_weight}Weight distribution of events using either the default
    \MEPSatNLO algorithm (red dashed) or the pilot scale strategy (blue solid)
    described in Sec.~\ref{sec:perfimprovs}.}
\end{figure}
The remaining question is whether the pilot run strategy adversely
affects the overall event weight. The large reduction in computing timing
observed in the last steps in Figs.~\ref{fig:perfimprovs:zjet:combined}
and~\ref{fig:perfimprovs:ttjet:combined} would then be reduced by the
unweighting efficiency in a second unweighting step that accounts for
the mismatch between the scale definitions.
Fig.~\ref{fig:pilot_weight} shows the weight distribution
of events after the complete simulation, i.e.\ including the matching
and merging procedure. We perform the analysis in partially unweighted mode,
which implies that the event weight can be modified by local
$K$-factors~\cite{Hoeche:2012yf}, and events are hence not fully unweighted.
Note that the distributions are presented on a logarithmic scale.
The average weights in the positive (negative) domain are 1.00 (-1.06) 
with a weight spread around 0.32 (0.52) when using the \MEPSatNLO algorithm 
and 1.03 (-1.12) with a weight spread around 0.40 (0.83) when using the 
pilot scale strategy for $pp \to e^+e^- + 0,1,2j$@NLO${}+3,4,5j$@LO.
For $pp \to t\bar{t} + 0,1j$@NLO${}+2,3,4j$@LO
the average weights in the positive (negative) domain are 1.02 (-1.23) 
with a weight spread around 0.65 (0.98) when using the \MEPSatNLO algorithm 
and 1.24 (-1.85) with a weight spread around 0.84 (1.59) when using the 
pilot scale strategy.
The efficiency of a second unweighting step that would make the event sample
unweighted can now be estimated as follows:
Determine the number of events to be generated. This corresponds to the area
under the curve, integrating from the top\footnote{In practice, one will
  need to account for the reduction in statistical power of the event sample
  due to negative weights: $(1-2f)^2$ for a negative weight fraction of $f$.}.
Find the weight of the right (left) edge of the area integrated over in the
positive (negative) half plane. The unweighting efficiency is the value of this
weight (i.e. the maximum weight at the given number of events) divided by
the average weight. Note that the average weight itself depends on the
number of events. For a large number of events and a sharply peaked weight
distribution, as in Fig.~\ref{fig:pilot_weight}, this effect can be ignored.
We find that the effective reduction in efficiency from using the pilot scale
approach is typically less than a factor of two if the target number of events is large.
The computing time reduction shown in the last steps in
Figs.~\ref{fig:perfimprovs:zjet:combined} and~\ref{fig:perfimprovs:ttjet:combined}
will effectively be reduced by this amount, but the usage of a pilot scale
is in most cases still beneficial.

\begin{figure}[tp]
  \includegraphics[width=0.32\textwidth]{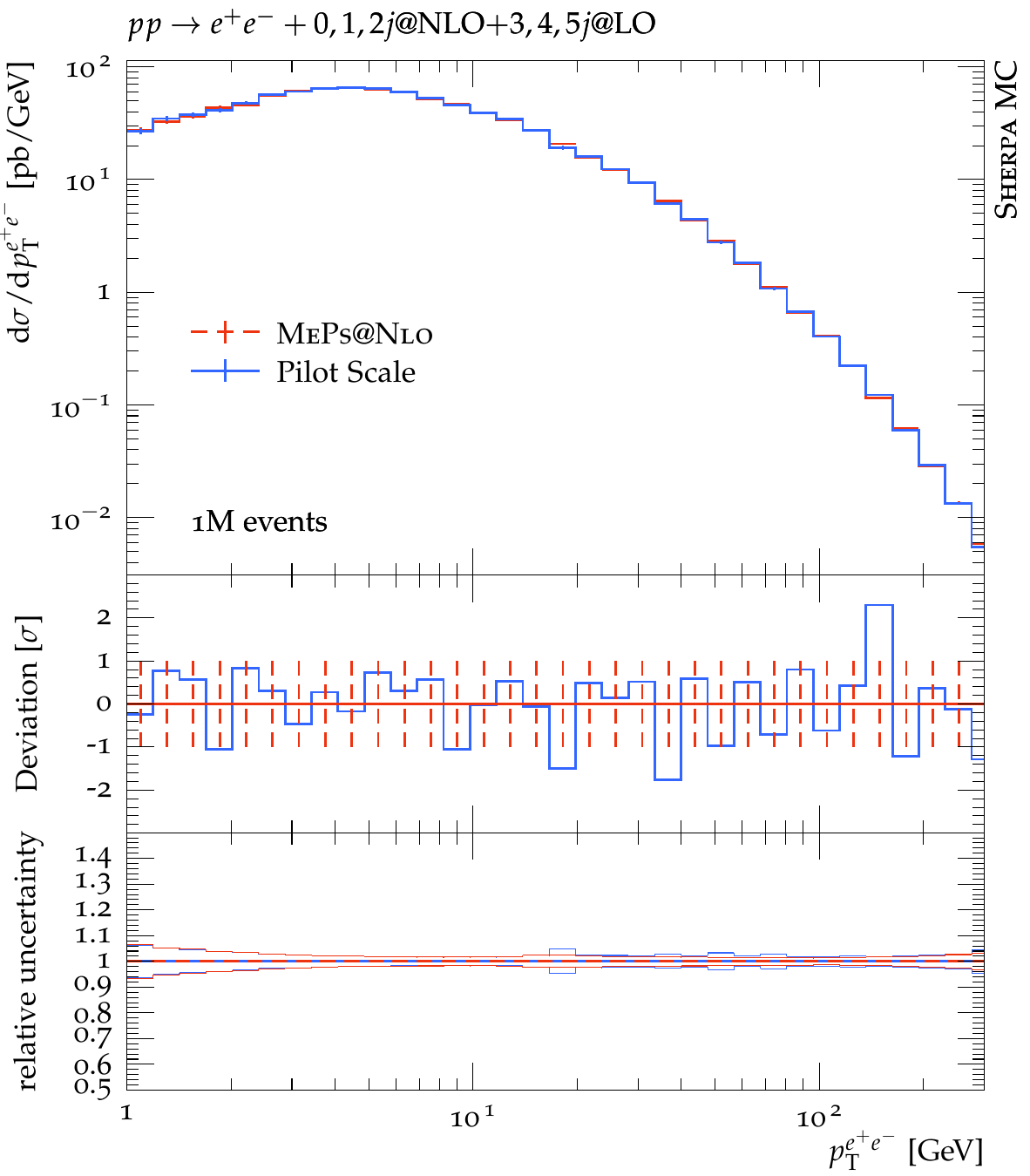}
  \hfill
  \includegraphics[width=0.32\textwidth]{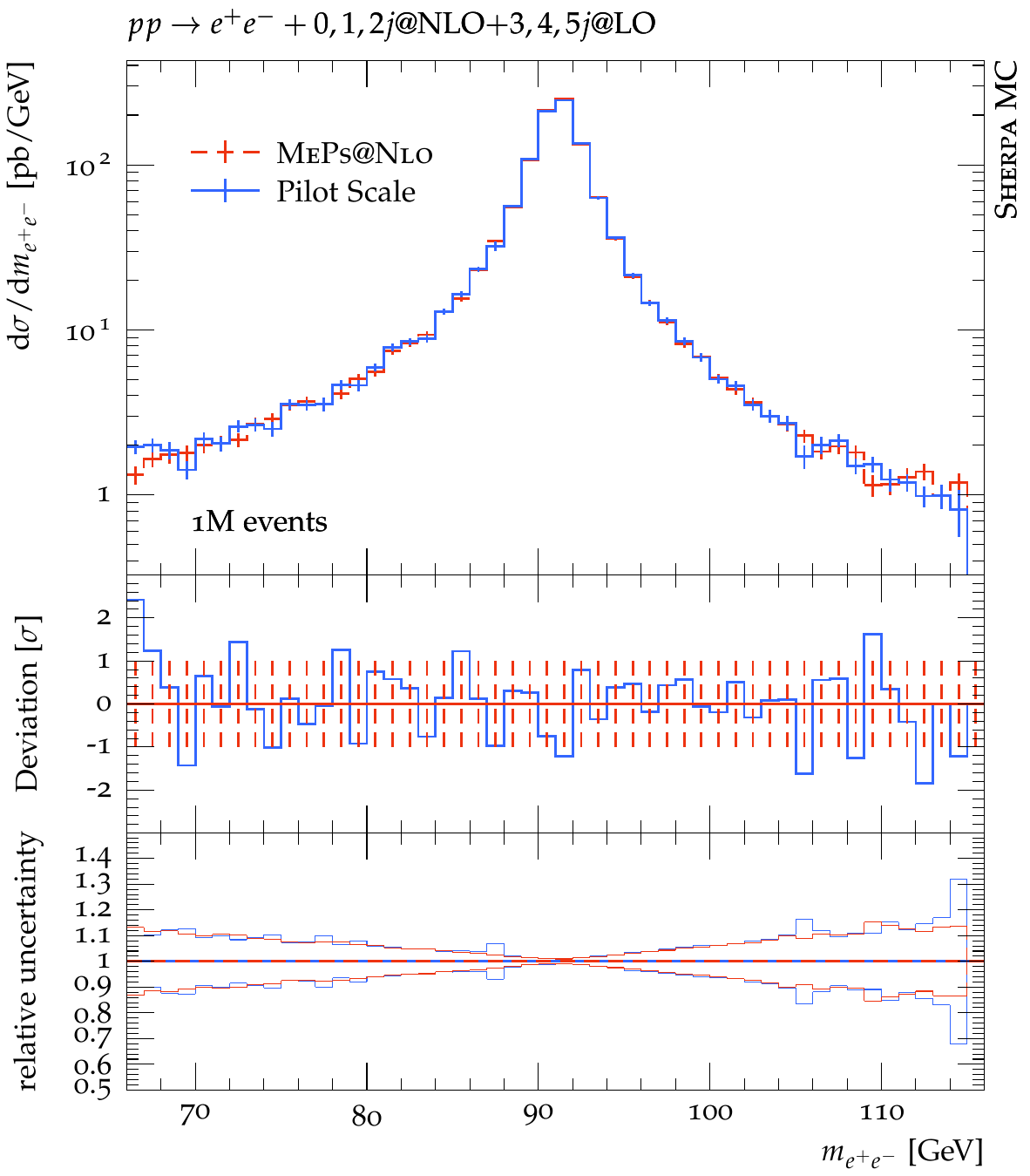}
  \hfill
  \includegraphics[width=0.32\textwidth]{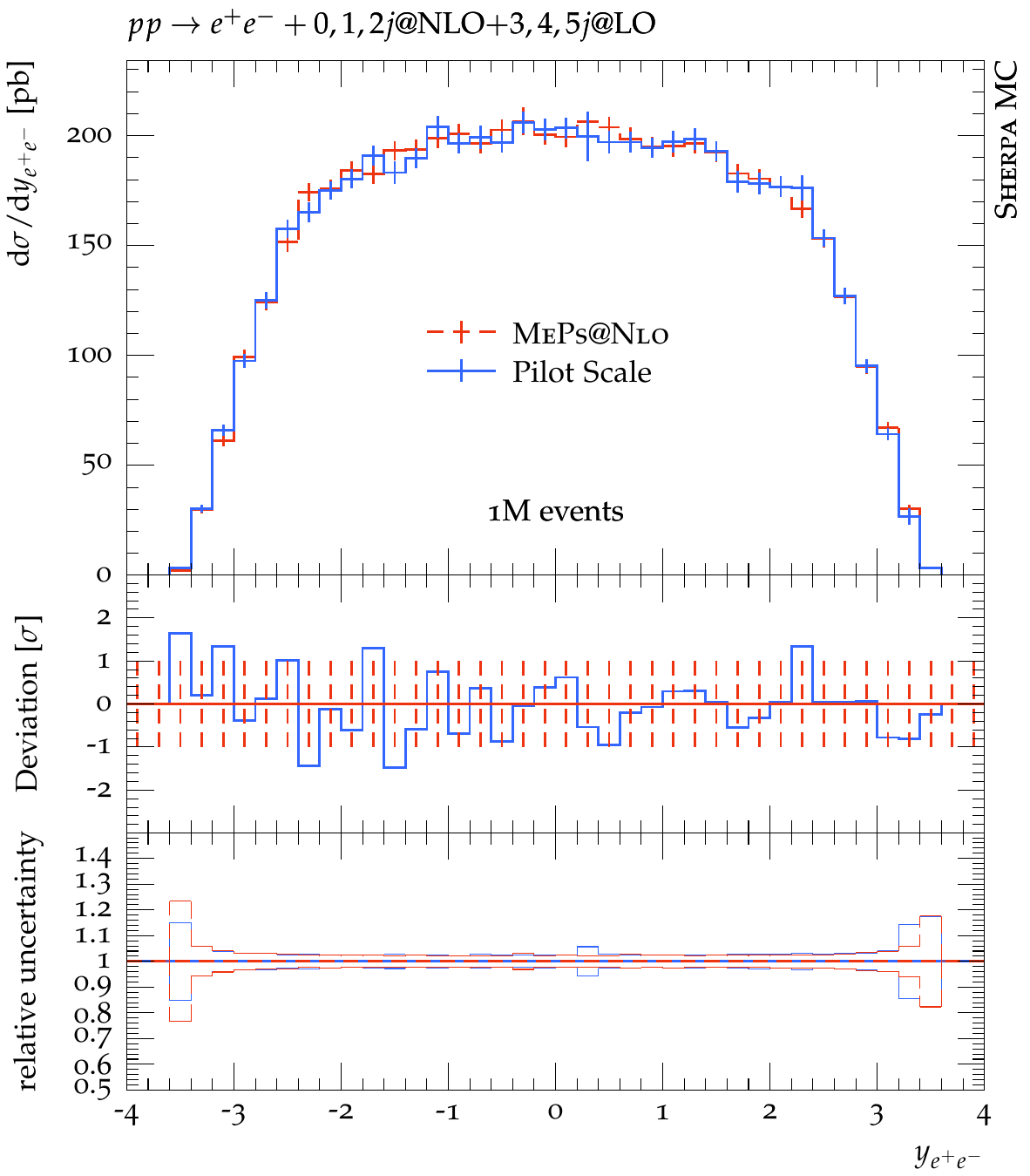}
  \caption{\label{fig:pilotweights:zjet:obs}%
    Predictions for the Born-level observables $m_{e^+e^-}$ and $y_{e^+e^-}$,
    and for the dilepton transverse momentum in comparison between the
    \MEPSatNLO algorithm (red dashed) or the pilot scale strategy (blue solid)
    described in Sec.~\ref{sec:perfimprovs}.
  }
\end{figure}

Finally, Fig.~\ref{fig:pilotweights:zjet:obs}
presents a cross-check between the \MEPSatNLO method and the new pilot run strategy
for actual $e^+e^-$ production physics observables
given our $pp \to e^+e^- + 0,1,2j$@NLO${}+3,4,5j$@LO setup.
We show distributions which can already be defined
at Born level ($m_{e^+e^-}$ and $y_{e^+e^-}$), as well as one observable which
probes genuine higher-order effects ($p_\mathrm{T}^{e^+e^-}$). We observe agreement
between the two scales at the statistical level, as well as MC uncertainties
of the same magnitude. This indicates that our new pilot run strategy will
be appropriate not only at the inclusive level, but also for fully differential
event simulation.
We have confirmed that the same conclusions are true
for the $pp \to t\bar{t} + 0,1j$@NLO${}+2,3,4j$@LO setup.

\begin{figure}[tp]
  \includegraphics[width=0.32\textwidth]{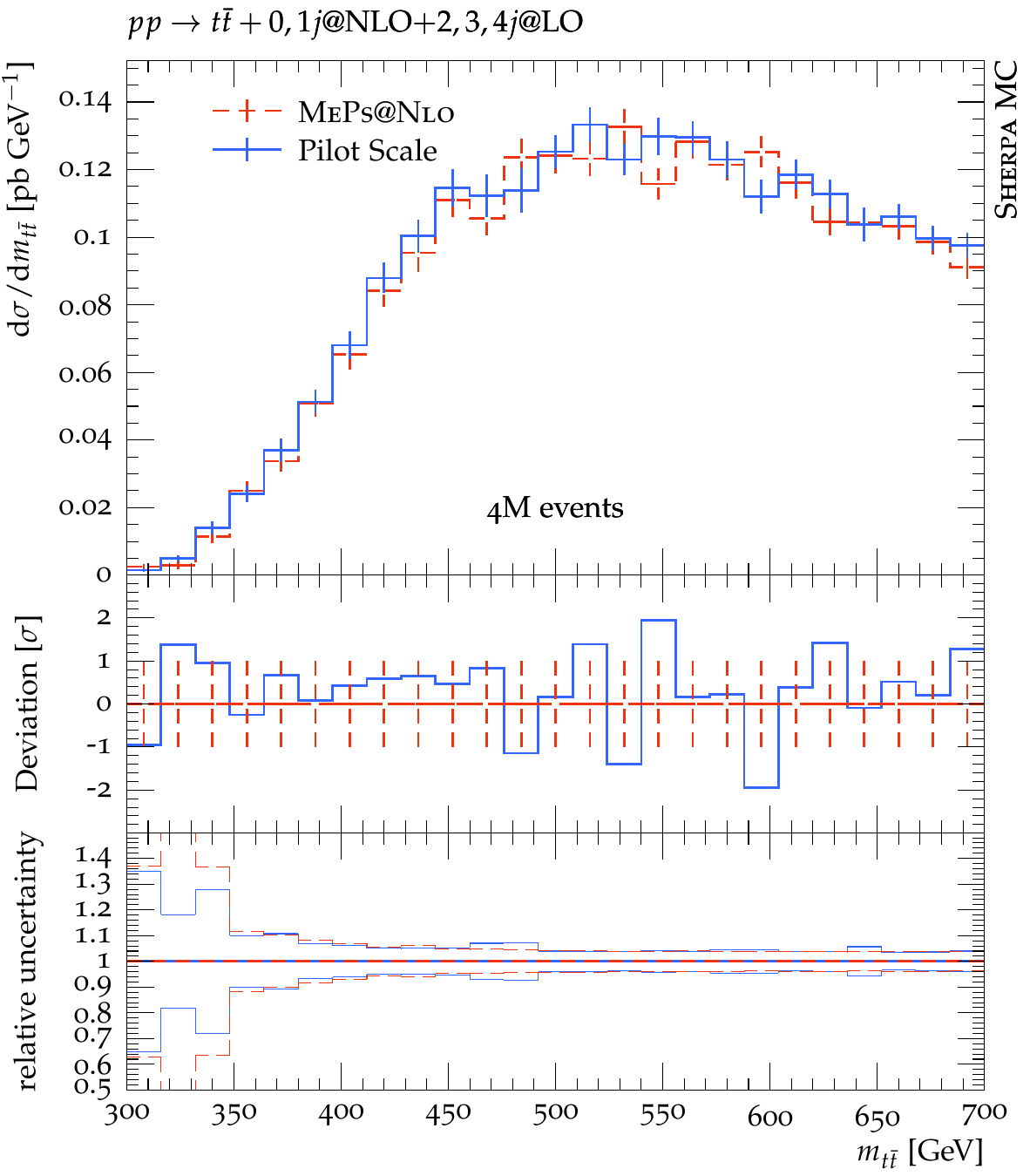}
  \hfill
  \includegraphics[width=0.32\textwidth]{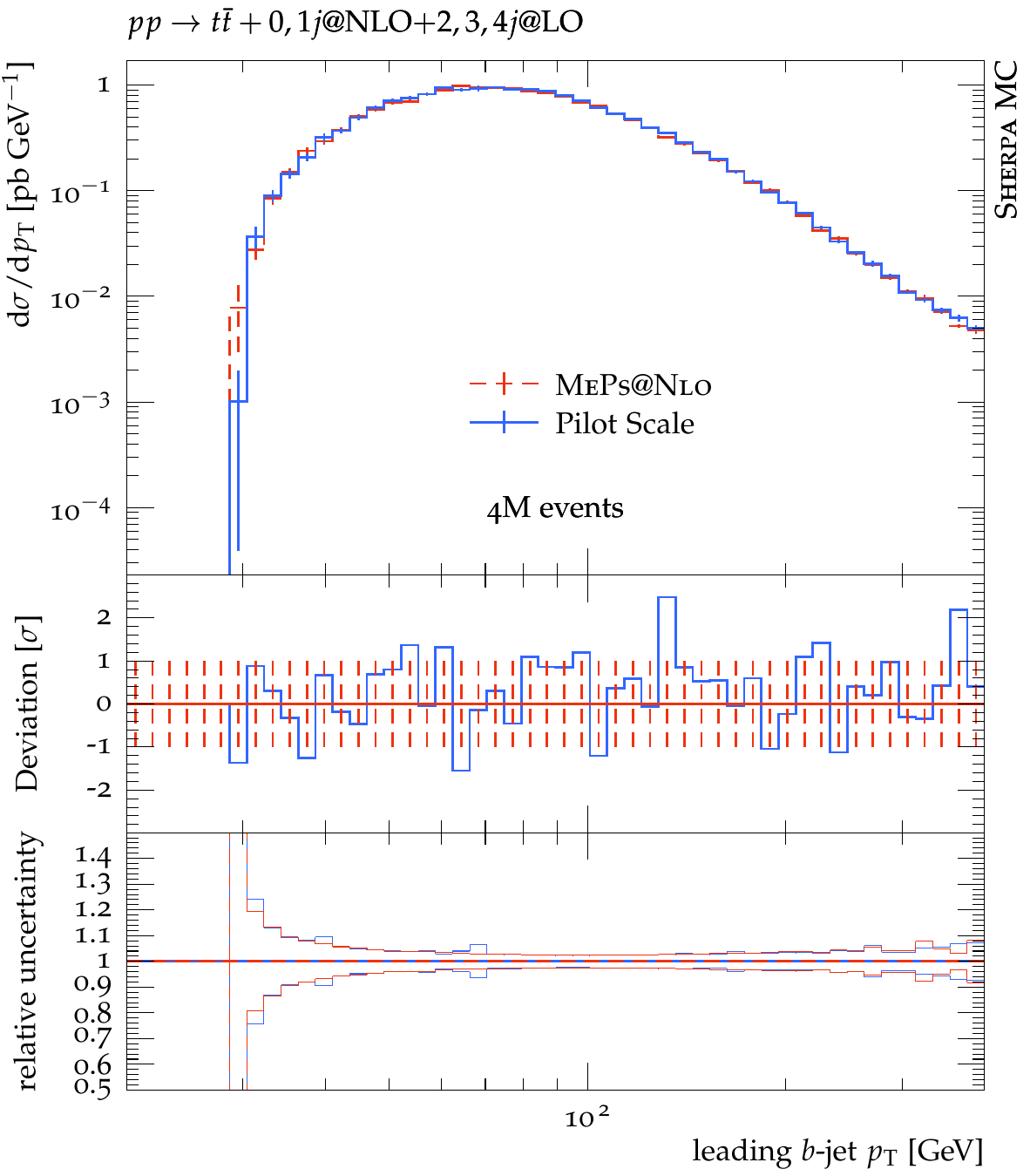}
  \hfill
  \includegraphics[width=0.32\textwidth]{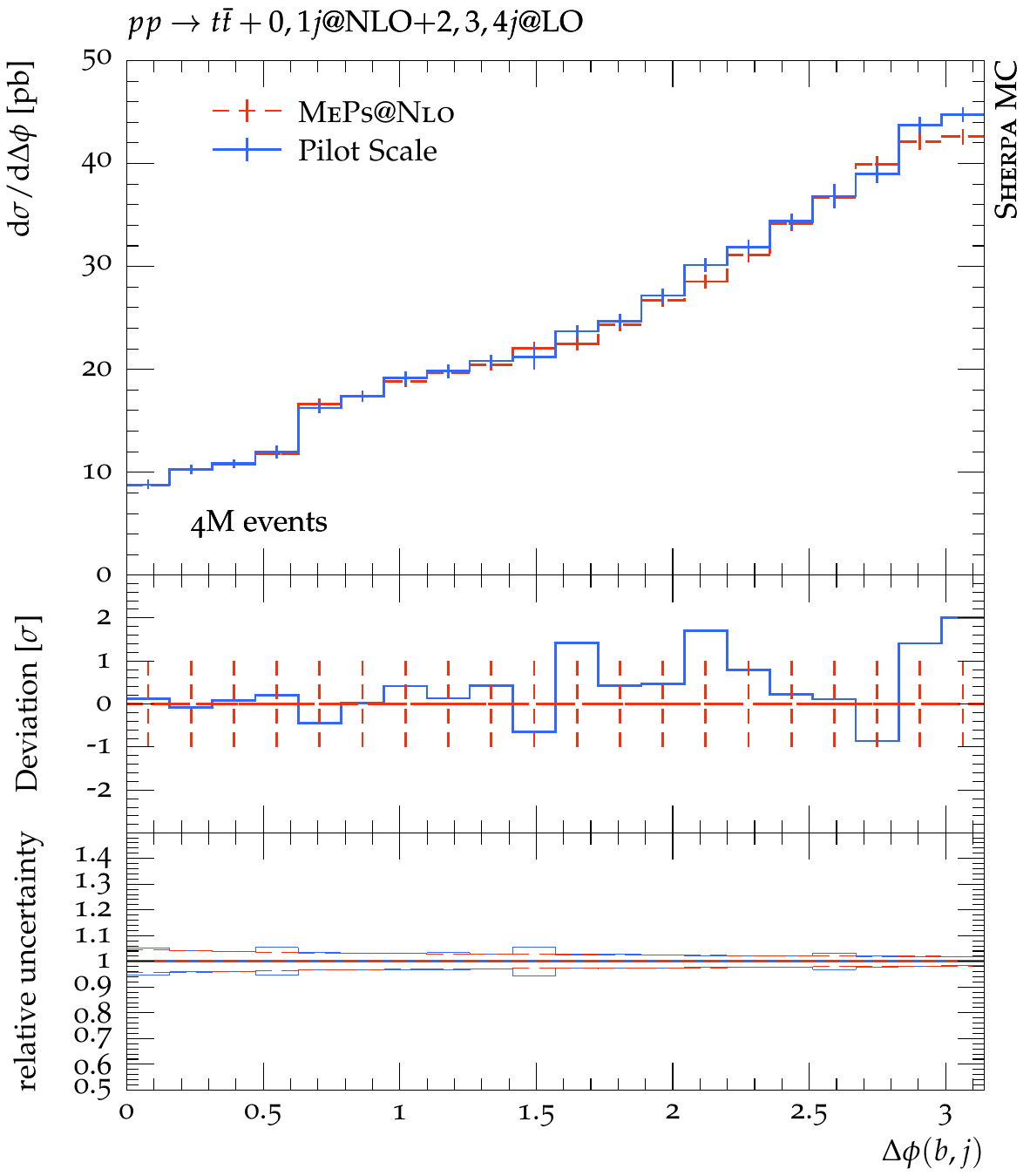}
  \caption{\label{fig:pilotweights:ttbar:obs}%
    Predictions for the Born-level observable $m_{t\bar{t}}$,
    for the leading $b$-jet transverse momentum and the azimuthal difference 
    between the leading $b$-jet and the leading light-flavour jet in comparison between the
    \MEPSatNLO algorithm (red dashed) or the pilot scale strategy (blue solid)
    described in Sec.~\ref{sec:perfimprovs}.
  }
\end{figure}



\section{Future performance improvements}
\label{sec:future}

We have shown, that for a large number of PDF variations, \LHAPDF still consumes a
significant portion of the computing time. While current realistic setups are
of roughly 100-200 variations, future analyses might require an ever increasing
number of variations and thus again an improved \LHAPDF and a better
PDF call alignment in \Sherpa.

The presented \LHAPDF performance improvements mostly depend on better caching
strategies. Future implementations might choose interpolators based on their
ability to precompute and store computations. For example, switching from a
2-step local polynomial interpolator to a ``proper'' bicubic interpolation would
allow to precompute all 16 coefficients of a third-order polynomial and only require
a matrix-vector multiplication at run-time.

In the context of \Sherpa in particular, with the increasing use of
multi-weights in the Monte Carlo event generation, the next step to even further
increase the caching of common computations would be to also cache the shared
computations of the error sets. This requires all the variations to be evaluated
at the same time, without changing the $(x,Q^2)$ point before moving to the next
one. This could be a further consideration if the number of variations increases
but requires a restructuring of the call pattern in \Sherpa.

However, currently for the realistic setups we presented, the majority of
computing-time is spent on phase space and matrix element computations which would
thus be the natural next step for performance improvements. In particular for
the high multiplicity matrix elements the generation of unweighted events suffers
from low unweighting efficiencies (which is also the reason why the pilot-run
yields such significant improvements).

A comparison between \Sherpa and \MCFM suggested that this computing time can be
further reduced~\cite{Campbell:2021vlt}: Firstly, Sherpa could make use of the
analytic tree-level matrix elements available in \MCFM. Secondly, the phase-space
integration strategy used by \MCFM could be adopted by \Sherpa in order
to increase efficiency.

In addition to these more traditional techniques, high multiplicity matrix elements
could be evaluated on GPUs, a path which has been charted in~\cite{Bothmann:2021nch}.
We expect significant improvements in this direction in the following
years~\cite{Campbell:2022qmc}.

Finally, the improvements presented in Sec.~\ref{sec:improvs:pilotrun} enable
Sherpa to be used for the processing of the HDF5 event files introduced
in~\cite{Hoche:2019flt}, both at leading and at next-to-leading order precision.
The corresponding technology is currently being implemented.

\section{Conclusion}
\label{sec:conclusion}
This manuscript discussed performance improvements of two major software packages
needed for event generation at the High-Luminosity LHC: \Sherpa and \LHAPDF.
We have presented multiple simple strategies to reduce the computing time
needed for unweighted event generation in these two packages, while maintaining
the formal precision of the computations. In combination, we achieve a reduction
of a factor 15 (7) in the computing requirements
for state-of-the-art $pp \to e^+e^- + 0,1,2j$@NLO${}+3,4,5j$@LO
($pp \to t\bar{t} + 0,1j$@NLO${}+2,3,4j$@LO) simulations at the LHC.
With this, we have achieved a major milestone set by the HSF event generator
working group and opened a path towards high-fidelity event simulation
in the HL-LHC era.
Our modifications are made publicly available for immediate use by the
LHC experiments.

\section*{Acknowledgments}
The authors would like to thank the Durham
Performance Analysis Workshop series \cite{basden_alastair_2021_5155503}
which provided an important forum to gain the necessary insights
during key phases of this work.
This research was supported by the Fermi National Accelerator Laboratory (Fermilab),
a U.S.\ Department of Energy, Office of Science, HEP User Facility.
Fermilab is managed by Fermi Research Alliance, LLC (FRA),
acting under Contract No. DE--AC02--07CH11359.
The work of S.H.\ was supported by the U.S. Department of Energy,
Office of Science, Office of Advanced Scientific Computing Research,
Scientific Discovery through Advanced Computing (SciDAC) program.
A.B., and M.S.\ are supported by the UK Science and Technology Facilities Council (STFC)
Consolidated Grant programmes ST/S000887/1 and\ ST/T001011/1, respectively.
A.B., I.C., C.G., and M.S.\ are further supported by the STFC SWIFT-HEP project (grant ST/V002627/1).
C.G.\ wishes to thank Edd Edmondson for technical support and providing exclusive use of a machine for benchmarking.
E.B. and M.K.\ acknowledge support from BMBF (contract 05H21MGCAB) and funding by
the Deutsche Forschungsgemeinschaft
(DFG, German Research Foundation) - project number 456104544.
M.K.\ would like to thank the MCnet collaboration for the opportunity to spend a short-term studentship at the University of Glasgow.
This work has received funding from the European Union's Horizon 2020 research and innovation programme as part of the
Marie Skłodowska-Curie Innovative Training Network MCnetITN3 (grant agreement no. 722104).
M.S.\ is supported by the Royal Society through a University
Research Fellowship (URF\textbackslash{}R1\textbackslash{}180549)
and an Enhancement Award (RGF\textbackslash{}EA\textbackslash{}181033,
CEC19\textbackslash{}100349, and RF\textbackslash{}ERE\textbackslash{}210397).
Our thanks to the MCnetITN3 network for enabling elements of
this project, and to Tushar Jain and the Google Summer of Code programme for initial
investigations of \LHAPDF performance.
I.C. and C.G. acknowledge the use of the UCL Myriad and Kathleen High Performance Computing Facilities,
and associated support services, in the completion of this work.

\appendix
\section{Run cards}
\label{app:runcards}

Listings~\ref{runcard-zee} and~\ref{runcard-ttbar} show the runcards
used in this study
for $Z+$jets and $t\bar{t}+$jets production, respectively.
Therein, we omit for brevity Standard Model parameter specifications,
which have no bearing on the findings of this study.

\begin{lrbox}{\myboxone}
\begin{minipage}{1.2\textwidth}
\begin{lstlisting}[
  caption={\Sherpa runcard for $Z+$jets production, in the 1000 variations setup.},
  label=runcard-zee,
  captionpos=b,
  escapeinside={/*!}{!*/},
%   escapebegin=\color{red}
]
(run){
  # Collider setup
  BEAM_1 2212; BEAM_ENERGY_1 6500.;
  BEAM_2 2212; BEAM_ENERGY_2 6500.;

  # PDF setup
  USE_PDF_ALPHAS 1;
  PDF_LIBRARY LHAPDFSherpa;
  PDF_SET NNPDF30_nnlo_as_0118;

  # ME generator settings
  ME_SIGNAL_GENERATOR Comix Amegic LOOPGEN PILOTLGEN;
  LOOPGEN:=OpenLoops;
  PILOTLGEN:=MCFM; # analytic virtual corrs., see Sec. /*!\ref{sec:improvs:analytics}!*/

  # Tags for steering the process setup
  NJET:=5; LJET:=2,3,4; QCUT:=20.;

  # Scale definitions
  SCALES PILOT{H_Tp2}{H_Tp2/sqr(2*max(1,N_FS-2))}; # simple pilot scale, see Sec. /*!\ref{sec:improvs:pilotscale}!*/
  PP_RS_SCALE VAR{H_Tp2/4};

  # Shower settings and neg. weight reduction
  NLO_SUBTRACTION_SCHEME 2;
  METS_BBAR_MODE 5;
  NLO_CSS_PSMODE 2;  # /*!$\langle\text{LC}\rangle$!*/-matching, see Sec. /*!\ref{sec:improvs:lcmatching}!*/
  PP_HPSMODE 0;
  OVERWEIGHT_THRESHOLD 10;

  # Variation setup
  SCALE_VARIATIONS 0.25,0.25 0.25,1. 1.,0.25 1.,1. 1.,4. 4.,1. 4.,4.;
  PDF_VARIATIONS NNPDF30_nnlo_as_0118_1000[all];
  ASSOCIATED_CONTRIBUTIONS_VARIATIONS EW EW|LO1;

  # Parton-shower reweighting setup
  CSS_REWEIGHT 1;
  REWEIGHT_SPLITTING_PDF_SCALES 1;
  REWEIGHT_SPLITTING_ALPHAS_SCALES 1;

  # Model setup
  EW_SCHEME 3;
  OL_PARAMETERS ew_renorm_scheme 1;
}(run)

(processes){
  Process 93 93 -> 11 -11 93{NJET};
  Order (*,2); CKKW sqr(QCUT/E_CMS);
  Enhance_Function VAR{max(pow(sqrt(H_T2)-PPerp(p[2])-PPerp(p[3]),2),PPerp2(p[2]+p[3]))/400.0} {3,4,5,6,7,8};
  Associated_Contributions EW|LO1 {LJET};
  NLO_QCD_Mode MC@NLO {LJET};
  ME_Generator Amegic {LJET};
  RS_ME_Generator Comix {LJET};
  Loop_Generator LOOPGEN {LJET};
  Pilot_Loop_Generator PILOTLGEN {LJET};
  Max_N_Quarks 4 {6,7,8};
  Max_Epsilon 0.01 {6,7,8};
  Integration_Error 0.99 {3,4,5,6,7,8};
  End process;
}(processes)

(selector){
  Mass 11 -11 40.0 E_CMS;
}(selector)
\end{lstlisting}
\end{minipage}
\end{lrbox}

\begin{figure*}[p]
  \centering
  \scalebox{0.8}{\usebox{\myboxone}}
\end{figure*}

\begin{lrbox}{\myboxtwo}
\begin{minipage}{1.2\textwidth}
\begin{lstlisting}[
  caption={\Sherpa runcard for $t\bar{t}+$jets production, in the 1000 variations setup.},
  label=runcard-ttbar,
  captionpos=b,
  escapeinside={/*!}{!*/},
%   escapebegin=\color{red}
]
(run){
  # Collider setup
  BEAM_1 2212; BEAM_ENERGY_1 6500.;
  BEAM_2 2212; BEAM_ENERGY_2 6500.;

  # PDF setup
  USE_PDF_ALPHAS 1;
  PDF_LIBRARY LHAPDFSherpa;
  PDF_SET NNPDF30_nnlo_as_0118;

  # ME generator settings
  ME_SIGNAL_GENERATOR Comix Amegic LOOPGEN PILOTLGEN;
  LOOPGEN:=OpenLoops;
  PILOTLGEN:=MCFM; # analytic virtual corrs., see Sec. /*!\ref{sec:improvs:analytics}!*/

  # Tags for steering the process setup
  NJET:=4; LJET:=2,3; QCUT:=30.;

  # Scale definitions
  SCALES PILOT{H_TM2}{H_TM2/sqr(2*max(1,N_FS-2))}; # simple pilot scale, see Sec. /*!\ref{sec:improvs:pilotscale}!*/
  CORE_SCALE QCD;
  EXCLUSIVE_CLUSTER_MODE 1;

  # Shower settings and neg. weight reduction
  NLO_SUBTRACTION_SCHEME 2;
  METS_BBAR_MODE 5;
  NLO_CSS_PSMODE 2;  # /*!$\langle\text{LC}\rangle$!*/-matching, see Sec. /*!\ref{sec:improvs:lcmatching}!*/
  PP_HPSMODE 0;
  OVERWEIGHT_THRESHOLD 10;

  # Variation setup
  SCALE_VARIATIONS 0.25,0.25 0.25,1. 1.,0.25 1.,1. 1.,4. 4.,1. 4.,4.;
  PDF_VARIATIONS NNPDF30_nnlo_as_0118_1000[all];
  ASSOCIATED_CONTRIBUTIONS_VARIATIONS EW EW|LO1;

  # Parton-shower reweighting setup
  CSS_REWEIGHT 1;
  REWEIGHT_SPLITTING_PDF_SCALES 1;
  REWEIGHT_SPLITTING_ALPHAS_SCALES 1;

  # Model setup
  EW_SCHEME 3;
  OL_PARAMETERS ew_renorm_scheme 1;

  # top and W decay setup
  HARD_DECAYS On;
  SOFT_SPIN_CORRELATIONS 1;

  HDH_STATUS[24,2,-1]    2; HDH_STATUS[24,4,-3]    2;
  HDH_STATUS[24,12,-11]  2; HDH_STATUS[24,14,-13]  2; HDH_STATUS[24,16,-15] 2;
  HDH_STATUS[-24,-2,1]   2; HDH_STATUS[-24,-4,3]   2;
  HDH_STATUS[-24,-12,11] 2; HDH_STATUS[-24,-14,13] 2; HDH_STATUS[-24,-16,15] 2;
  STABLE[24] 0; STABLE[6] 0; WIDTH[6] 0;
}(run)

(processes){
  Process 93 93 -> 6 -6 93{NJET};
  Order (*,0); CKKW sqr(QCUT/E_CMS);
  Enhance_Function VAR{pow(max(sqrt(H_T2)-PPerp(p[2])-PPerp(p[3]),(PPerp(p[2])+PPerp(p[3]))/2)/30.0,2)} {3,4,5,6}
  Associated_Contributions EW|LO1 {LJET};
  NLO_QCD_Mode MC@NLO {LJET};
  ME_Generator Amegic {LJET};
  RS_ME_Generator Comix {LJET};
  Loop_Generator LOOPGEN {LJET};
  Pilot_Loop_Generator PILOTLGEN {2};
  End process;
}(processes)
\end{lstlisting}
\end{minipage}
\end{lrbox}

\begin{figure*}[p]
  \centering
  \scalebox{0.8}{\usebox{\myboxtwo}}
\end{figure*}

\bibliographystyle{amsunsrt_mod}
\bibliography{journal}
\end{document}